\documentclass[a4paper,11pt]{article}
\pdfoutput=1 % if your are submitting a pdflatex (i.e. if you have
\usepackage{jheppub} % for details on the use of the package, please
\usepackage[T1]{fontenc} % if needed
\usepackage{comment}

\title{\boldmath Spinor-Helicity Formalism for Massless Fields in AdS$_4$ II: Potentials}

\author[a]{Balakrishnan Nagaraj}
\author[b,c]{and Dmitry Ponomarev}

\affiliation[a]{George P. and Cynthia W. Mitchell Institute for Fundamental Physics and Astronomy,\\
 Texas A\&M University, University Drive,  College Station, TX 77843, USA}
\affiliation[b]{Institute for Theoretical and Mathematical Physics,\\
Lomonosov Moscow State University, Lomonosovsky avenue, Moscow, 119991, Russia}
\affiliation[c]{I.E. Tamm Theory Department, Lebedev Physical Institute,\\
 Leninsky avenue, Moscow, 119991, Russia}
\emailAdd{nbala1090@gmail.com}
\emailAdd{ponomarev@lpi.ru}

\abstract{In a recent letter we suggested a natural  generalization of the flat-space spinor-helicity formalism in four dimensions to anti-de Sitter space. In the present paper we give  some technical details that were left implicit previously. For lower-spin fields we also derive potentials associated with the previously found plane wave solutions for  field strengths. We then employ these potentials to evaluate some three-point amplitudes. This analysis illustrates a typical computation of an amplitude without internal lines in our formalism.}

\begin{document} 
\maketitle
\flushbottom

\section{Introduction}

In recent years significant progress was achieved in amplitudes' computations as well as in understanding of various hidden structures underlying them. This is especially true for theories of massless particles in four dimensions. For these theories one can choose convenient kinematic variables that lead to what is known as the spinor-helicity formalism. This formalism allows to compute amplitudes efficiently and produces them in an extremely compact form. This is typically illustrated by the  Parke-Taylor formula \cite{Parke:1986gb}, which gives a single-term expression for a tree-level MHV $n$-point amplitude in the Yang-Mills theory. The spinor-helicity formalism also fits together nicely with other techniques used for amplitudes' computations. For review on modern amplitude methods and on the spinor-helicity formalism, see \cite{Dixon:1996wi,Elvang:2013cua,Dixon:2013uaa}.
The success of the spinor-helicity formalism for theories of massless particles in four dimensions motivated its various extensions --- to other dimensions \cite{Cheung:2009dc,CaronHuot:2010rj,Boels:2012ie,Bandos:2016tsm,Bandos:2017eof}
and to massive fields \cite{Conde:2016vxs,Conde:2016izb,Arkani-Hamed:2017jhn,Jha:2018hag}.

Another line of research that lead to important developments in recent years is the AdS/CFT correspondence. It is a conjectured duality between gravitational theories in AdS space and conformal theories on its boundary \cite{Maldacena:1997re,Gubser:1998bc,Witten:1998qj}. The AdS/CFT correspondence provides us with new tools to address important problems of quantum gravity and strongly coupled systems. On the AdS side perturbative observables are computed by  Witten diagrams, which can be regarded as the AdS counterpart of flat scattering amplitudes. These diagrams can be expressed in different representations: in terms of boundary coordinates that label external lines, in terms of the associated Fourier or Mellin space variables or presented in the form of the conformal block decomposition, see \cite{Freedman:1998tz,Liu:1998ty,DHoker:1999kzh,Heemskerk:2009pn,Penedones:2010ue,Paulos:2011ie,Fitzpatrick:2011dm,Rastelli:2016nze,Aharony:2016dwx,Aprile:2017bgs,Yuan:2018qva,Albayrak:2018tam} for a far from complete list of references. Each of these representations has its own virtues and for each of them major progress was achieved in recent years. In particular, more efficient methods of computing  Witten diagrams were developed,  relations between the analytic structure of amplitudes and types of bulk processes were understood, it was found how take the flat-space limit of Witten diagrams, thus, reproducing the associated flat scattering amplitudes. Moreover, these results can be extended to dS space producing de Sitter space correlators, which, in turn, are closely related to inflationary correlators, see e.g. \cite{Maldacena:2011nz,Arkani-Hamed:2015bza,Lee:2016vti,Arkani-Hamed:2017fdk,Arkani-Hamed:2018kmz}. Despite these successes, the aforementioned approaches typically require one to deal with complicated expressions, that involve various special functions. Moreover, the analysis further complicates for spinning fields due to proliferation of tensor indices. This begs the question: is there any natural generalization of the spinor-helicity formalism to AdS, which allows to deal with amplitudes of massless fields as efficiently as in flat space?

Additional motivation to address this question is related to higher-spin theories. It was discovered recently \cite{Bengtsson:2014qza,Conde:2016izb}  that the spinor-helicity formalism allows to construct additional consistent cubic amplitudes compared to those available within the framework that employs Lorentz tensors. This observation is based on the comparison of two classifications available in both approaches. Moreover, it turns out that the associated cubic vertices are crucial for consistency of higher-spin interactions in flat space \cite{Metsaev:1991mt,Metsaev:1991nb}\footnote{These results were obtained employing the light-cone deformation procedure, which is closely connected to the spinor-helicity formalism \cite{Ananth:2012un,Bengtsson:2016jfk,Ponomarev:2016cwi}.}. In particular, these are present in chiral higher-spin theories \cite{Ponomarev:2016lrm,Ponomarev:2017nrr,Skvortsov:2018jea} ---  cubic theories of massless higher-spin fields, which are consistent to all orders in interactions \cite{Ponomarev:2016lrm}, see also \cite{Devchand:1996gv} for a related earlier result.

At the same time, the AdS/CFT correspondence implies the existence of higher-spin theories in AdS space --- holographic duals of free $O(N)$ vector models and similar simple theories \cite{Klebanov:2002ja,Sezgin:2002rt}\footnote{An independent bulk formulation of higher-spin theories was proposed in \cite{Vasiliev:1990en,Vasiliev:1992av}. For recent discussions of this approach, see \cite{Giombi:2009wh,Boulanger:2015ova,Didenko:2018fgx,Didenko:2019xzz}.}. One may wonder how these theories are related to chiral theories in flat space. To be able to answer this question, it is important to develop an approach, that would bridge the gap between the light-cone formulation of chiral theories in flat space and the usual covariant language in AdS space, which is typically used in holography. This approach can then be used to generalize chiral higher-spin theories to AdS space and study their place in the holographic duality. It is also interesting to understand what happens with additional vertices in AdS space. Their presence may play an important role not only in higher-spin theories, but in a wider context. For example, it would be interesting to understand whether the associated three-point correlators can appear in conformal field theories or, more generally, whether the spinor-helicity representation can facilitate the analysis of the crossing equations that involve spinning operators.

In a recent letter \cite{Nagaraj:2018nxq} we suggested a natural generalization of the spinor-helicity formalism to AdS${}_4$\footnote{A different version of the spinor-helicity formalism in AdS${}_4$ was suggested in \cite{Maldacena:2011nz}. Though this formalism has its own virtues, in some aspects it departs from the spinor-helicity formalism in flat space. In particular, it does not make the Lorentz symmetry manifest, instead, employing $so(3)$-spinors. More recent related result  can be found in \cite{David:2019mos}.}. This approach is based on the standard realization of the isometry algebra $so(3,2)$ of AdS${}_4$ in terms of differential operators acting on $sl(2,\mathbb{C})$ spinors. By employing this representation, we first found the AdS counterpart of plane waves for field strengths. These solutions were then used to compute simplest amplitudes. Next, by employing the symmetry arguments similar to those used in \cite{Benincasa:2007xk} in flat space, we classified three-point amplitudes of spinning fields in AdS${}_4$. As was expected, the spinor-helicity approach allows to construct amplitudes, that cannot be represented in terms of Lorentz tensors. This result is consistent with a recent analysis in the light-cone gauge \cite{Metsaev:2018xip,Skvortsov:2018uru}.

In the present paper we give technical details that were left implicit in \cite{Nagaraj:2018nxq}. Moreover, we expand these results in one important way. Namely, we show how our previous analysis can be extended to include the potentials of gauge fields. First, we find the plane wave solutions in terms of  potentials. Unlike plane wave solutions for field strengths, these cannot be obtained simply by applying Weyl transformations to flat space solutions, because the description of massless fields in four dimensions in terms of potentials is not conformally invariant for spin greater than one. We work out in detail  spin-$\frac{3}{2}$ and spin-$2$ cases and then comment on potentials of any spin. Next, we use these potentials to evaluate simple three-point amplitudes. Unlike amplitudes we computed previously, for which essential simplification occurred due to conformal invariance of the associated vertices or due to  the possibility to express them in terms of field strengths, in the present paper we deal with the cases, in which no such simplifications occur. These examples, thus, illustrate a genuine computation of a three-point amplitude using the spinor-helicity formalism in AdS${}_4$.

The rest of the paper is organized as follows. In section \ref{sec:2} we review  the ingredients of the spinor-helicity formalism in flat space, that will be generalized to AdS space later. Then, in section \ref{sec:3} we review the twisted adjoint representation --- a representation for massless fields in AdS${}_4$ in terms of differential operators, that act on $sl(2,\mathbb{C})$ spinors. In the following section we introduce  the necessary objects of the AdS background geometry. Then, in section \ref{sec:4} we derive plane wave solutions for field strengths. Next, in section \ref{sec:5} we discuss how the spinor-helicity formalism in AdS  space should be extended to include potentials and derive the associated solutions for lower-spin cases. In section \ref{sec:6} we use previously derived plane waves to compute amplitudes by the direct evaluation of bulk integrals. In the next section we classify  three-point amplitudes employing symmetry considerations. In section \ref{sec:8} we discuss how different amplitudes can be generated one from another by applying helicity-changing operators. Finally, we conclude in section \ref{sec:9} as well as discuss further open problems. The paper has a number of appendices, in which we collect our notations and present various technical results.

\section{Spinor-Helicity Representation in Flat Space}
\label{sec:2}

In this section we review some aspects of the spinor-helicity formalism in flat space, that will be later extended to AdS space. More details can be found in \cite{Dixon:1996wi,Elvang:2013cua,Dixon:2013uaa}.

Massless representation of the four-dimensional Poincare algebra can be realized as 
\begin{equation}
\label{22oct1}
\begin{split}
\mathcal{J}_{\alpha \beta}&=i \left(\lambda_{\alpha}\frac{\partial}{\partial\lambda^{\beta}}+\lambda_{\beta}\frac{\partial}{\partial\lambda^{\alpha}}\right),\\
\mathcal{\bar{J}}_{\dot{\alpha}\dot{\beta}}&=i\left(\bar{\lambda}_{\dot{\alpha}}\frac{\partial}{\partial\bar{\lambda}^{\dot{\beta}}}+\bar{\lambda}_{\dot{\beta}}\frac{\partial}{\partial\bar{\lambda}^{\dot{\alpha}}}\right),\\
\mathcal{P}_{\alpha\dot{\alpha}}&=\lambda_{\alpha}\bar{\lambda}_{\dot{\alpha}},\\
\end{split}
\end{equation}
where $\lambda_{\alpha}$ is an $sl(2,\mathbb{C})$ spinor and $\bar\lambda_{\dot\alpha}$ is its complex conjugate. These spinors are related to real massless momenta by the standard vector-spinor dictionary
\begin{equation}
\label{22oct2}
p^2=0 \quad \Leftrightarrow \quad p_a=-\frac{1}{2}(\sigma_a)^{\dot\alpha\alpha}\lambda_\alpha\bar\lambda_{\dot\alpha},
\end{equation}
where $\sigma_\mu$ are the Pauli matrices. To make three-point amplitudes non-vanishing, one allows momenta to be complex. 
In this case $\lambda_\alpha$ and $\bar\lambda_{\dot\alpha}$ are independent.
Below we will use the vector-spinor dictionary quite extensively. 
A brief review of this dictionary and related conventions are given in appendix \ref{App:A}.

It is not hard to see that the helicity operator
\begin{equation}
\label{22oct3}
    H\equiv\frac{1}{2}\left(\Bar{\lambda}_{\dot{\alpha}}\frac{\partial}{\partial \Bar{\lambda}_{\dot{\alpha}}}-\lambda_{\alpha}\frac{\partial}{\partial \lambda_{\alpha}}\right)
\end{equation}
commutes with the generators of the Poincare algebra (\ref{22oct1}). This implies that the representation space -- that is the space of functions $f(\lambda,\bar\lambda)$ of $\lambda$ and $\bar\lambda$ on which operators (\ref{22oct1}) act -- can be split into a direct sum of representations with a definite value of $H$. These representations turn out to be irreducible. The value of  operator $H$ on these representations gives their helicity
\begin{equation}
\label{22oct4}
H=h,\qquad 2h\in \mathbb{Z}.
\end{equation}

\subsection{Plane wave solutions}
\label{sec:2.1}
To give these representations a space-time interpretation one introduces plane waves. These can be regarded as intertwining kernels between the spinor-helicity (\ref{22oct1}) and space-time representations of the Poincare algebra
\begin{equation}
\label{22oct5}
\begin{split}
{\cal P}_a=-i\frac{\partial}{\partial x^a}, \qquad {\cal J}_{ab}=-i\left(x_a\frac{\partial}{\partial x^b}-x_b\frac{\partial}{\partial x^a}\right),
\end{split}
\end{equation}
the latter being generated by the algebra of Killing vector fields of the Minkowski space-time. 

To start, we consider plane waves for field strengths. These are gauge invariant and simpler to find. For a particular helicity $2h=n\ge 0$, we solve for field strength's plane waves in the form
\begin{equation}
\label{22oct6}
F_{\dot\alpha_1\dots\dot\alpha_n}=\bar\lambda_{\dot\alpha_1}\dots \bar\lambda_{\dot\alpha_n}f(x,\lambda,\bar\lambda).
\end{equation}
The prefactor on the right hand side of (\ref{22oct6}) consisting of the $n$-fold product of spinors $\bar\lambda$ was introduced to saturate the homogeneity degree of $F$ in $\lambda$ and $\bar\lambda$ as required by (\ref{22oct3}), (\ref{22oct4}). Lorentz covariance requires that $f$ may only depend on combinations of $\lambda$, $\bar\lambda$ and $x$ with all indices contracted covariantly. In other words,
\begin{equation}
\label{22oct7}
f(x,\lambda,\bar\lambda) = d(a,b), \qquad a\equiv \lambda_\alpha\bar\lambda_{\dot\alpha} x^{\alpha\dot\alpha}, \qquad
b\equiv x_{\alpha\dot\alpha} x^{\dot\alpha\alpha},
\end{equation}
where $d$ is a new unknown function.
Finally, we require that the action of translations on plane waves agrees in the spinor-helicity and the space-time representations
\begin{equation}
\label{22oct8}
-i(\sigma^a)_{\beta\dot\beta}\frac{\partial}{\partial x^a} \bar\lambda_{\dot\alpha_1}\dots \bar\lambda_{\dot\alpha_n}d(a,b)
= \lambda_\beta \bar\lambda_{\dot\beta} \bar\lambda_{\dot\alpha_1}\dots \bar\lambda_{\dot\alpha_n} d(a,b).
\end{equation}
This leads to the familiar formula
\begin{equation}
\label{22oct9}
F_{\dot\alpha_1\dots\dot\alpha_n}=\bar\lambda_{\dot\alpha_1}\dots \bar\lambda_{\dot\alpha_n} e^{-\frac{i}{2}x^{\dot\alpha\alpha}\lambda_\alpha\bar\lambda_{\dot\alpha}}.
\end{equation}
Plane wave solutions for field strengths with negative helicities can be derived analogously.

Once field strengths are known, one can find the potentials. For bosonic fields these are related by \cite{deWit:1979sib,Francia:2002aa,Sorokin:2004ie}
\begin{equation}
\label{22oct10}
F_{a_1b_1,\dots,a_hb_h}=\partial_{a_1}\dots \partial_{a_h} \varphi_{b_1\dots b_h} +\dots,
\end{equation}
where $h_{a_1\dots a_h}$ is a totally symmetric tensor and $\dots$ denotes $2h-1$ terms to be added to make the expression antisymmetric in each pair of indices $\{a_i,b_i\}$. For a fixed field strength (\ref{22oct10}) defines the potential up to gauge transformations
\begin{equation}
\label{22oct11}
\delta \varphi_{a_1\dots a_h}=\partial_{a_1} \xi_{a_2\dots a_h}+\dots,
\end{equation}
where $\xi$ is totally symmetric and  $\dots$ denotes terms that make the right hand side totally symmetric. 

In the spinor-helicity formalism, for a field strength given by (\ref{22oct9}) one solves (\ref{22oct10}) for the helicity $h\ge 0$ mode as
\begin{equation}
\label{22oct12}
\varphi_{a_1\dots a_h} =\varepsilon^+_{a_1}\dots  \varepsilon^+_{a_h}e^{-\frac{i}{2}x^{\dot\alpha\alpha}\lambda_\alpha\bar\lambda_{\dot\alpha}},
\end{equation}
where $\varepsilon^+$ is a polarization vector 
\begin{equation}
\label{22oct13}
\varepsilon^+_a\equiv -\frac{i}{2}\frac{(\sigma_a)^{\dot\alpha\alpha}\mu_\alpha\bar\lambda_{\dot\alpha}}{\mu^\beta \lambda_\beta},
\end{equation}
defined in terms of an auxiliary spinor $\mu$. It is not hard to show that changes of $\mu$ correspond to gauge transformations. 

The potential (\ref{22oct12}) has a list of remarkable properties: it is traceless
\begin{equation}
\label{22oct14}
\eta^{ab}\varepsilon_a^+ \varepsilon_b^+=0 \qquad \Rightarrow \qquad \varphi^{b}{}_{b a_3\dots a_h}=0,
\end{equation}
divergence-free
\begin{equation}
\label{22oct15}
\partial^b\varphi_{b a_2\dots a_h}=0
\end{equation}
and obeys
\begin{equation}
\label{22oct16}
q^{b}\varphi_{ba_2\dots a_h}=0, \qquad q_a\equiv -\frac{1}{2}(\sigma_a)^{\dot\alpha\alpha}\mu_\alpha\bar\mu_{\dot\alpha}.
\end{equation}
Considering that $q$ is null, (\ref{22oct15}) can be regarded as the generalized light-cone gauge condition with the only difference that in the spinor-helicity formalism we are free to change $q$ arbitrarily. The potential (\ref{22oct12}), in fact, satisfies
\begin{equation}
\label{19nov1}
\mu^{\beta}(\sigma^b)_{\beta\dot\beta} \varphi_{ba_2\dot a_h}=0,
\end{equation}
which is a stronger version of (\ref{22oct16}).

For fermionic field the field strength is defined by
\begin{equation}
\label{22oct17}
F_{\dot\alpha|a_1b_1,\dots,a_{h-1}b_{h-1}}=\partial_{a_1} \dots \partial_{a_{n-1}}\varphi_{\dot\alpha|b_1\dots b_{h-1}} +\dots,
\end{equation}
where $\dots$ make the right hand side antisymmetric in $\{a_i,b_i\}$ and $\varphi$ is a totally symmetric spin-tensor in vector indices. Gauge transformations then read
\begin{equation}
\label{22oct18}
\delta\varphi_{\dot\alpha|a_1\dots a_{h-1}}= \partial_{a_1} \xi_{\dot\alpha| a_2\dots a_{h-1}}+\dots,
\end{equation}
where $\xi$ is totally symmetric in vector indices and $\dots$ make the right hand side totally symmetric. Finally, the plane wave solution (\ref{22oct9}) in terms of the potential reads
\begin{equation}
\label{22oct19}
\varphi_{\dot\alpha|a_1\dots a_{h-1}} =\bar\lambda_{\dot\alpha}\varepsilon^+_{a_1}\dots  \varepsilon^+_{a_{h-1}}e^{-\frac{i}{2}x^{\dot\alpha\alpha}\lambda_\alpha\bar\lambda_{\dot\alpha}}.
\end{equation}
Besides being traceless on vector indices, this potential is also $\sigma$-traceless in the sense that
\begin{equation}
\label{22oct19x1}
(\sigma^{b})^{\dot\alpha\alpha}\varphi_{\dot\alpha|b a_2\dots a_{h-1}} =0.
\end{equation}
It also satisfies conditions analogous to (\ref{22oct15})-(\ref{19nov1}).

For negative helicity fields one uses the complex conjugate of (\ref{22oct13}) as a polarization vector.

\subsection{Amplitudes from space-time integrals}
\label{sec:2.2}

Using the standard Feynman rules, with the external lines represented by plane-wave solutions we reviewed above, we can obtain the spinor-helicity representation of any amplitude.  Below we illustrate this with a simple example of a cubic vertex
\begin{equation}
\label{22oct20}
S_3=\int d^4 x\psi_{\mu|\alpha}(\partial^\mu \chi^{\alpha}\phi - \chi^\alpha \partial^\mu\phi)+\text{c.c.},
\end{equation}
where c.c. refers to the complex conjugate term, $\psi$, $\chi$ and $\phi$ are massless spin-$\frac{3}{2}$, spin-$\frac{1}{2}$ and spin-0 fields respectively.

It is not hard to see that (\ref{22oct20}) is invariant with respect to spin 3/2 gauge transformations provided free equations of motion are taken into account
\begin{equation}
\label{22oct21}
\Box \chi^\alpha \approx 0, \qquad \Box \phi\approx 0.
\end{equation}
This, in turn, implies that (\ref{22oct20}) can be made gauge invariant up to higher orders in fields, once $\phi$ and $\chi$ transform appropriately with the gauge transformations of $\psi$. In other words, (\ref{22oct21}) is a consistent vertex to the leading order in interactions. 

By substituting the plane-wave solutions into the first term in (\ref{22oct20}) we find the amplitude
\begin{equation}
\label{22oct22}
\begin{split}
{\cal A}_3\left(-\frac{3}{2},-\frac{1}{2},0\right)&= \frac{1}{2} \int d^4x e^{ip\cdot x} \frac{\langle 12 \rangle}{[1\mu]}\left(\langle 12\rangle [\mu 2]-\langle 13\rangle [\mu 3]   \right)\\
&\qquad =
\frac{(2\pi)^4}{2}  \frac{\langle 12 \rangle}{[1\mu]}\left(\langle 12\rangle [\mu 2]-\langle 13\rangle [\mu 3]   \right)\delta^4(p),
\end{split}
\end{equation}
where $p\equiv p_1+p_2+p_3$ is the total momentum\footnote{Though, the momentum-conserving delta function is not, usually, included in the definition of the amplitude, we do the opposite to facilitate the comparison with the AdS case.}.
Next, we would like to eliminate $\mu$-dependence, to make sure that the amplitude is gauge invariant. To this end we manipulate the first term in brackets as 
\begin{equation}
\label{22oct23}
\frac{\langle 12\rangle^2 [\mu 2]}{[1\mu]} \delta^4(p)=\frac{\langle 12\rangle^2 [\mu 2]\langle 23\rangle}{[1\mu]\langle 23\rangle}\delta^4(p)=-\frac{\langle 12\rangle^2 [\mu 1]\langle 13\rangle}{[1\mu]\langle 23\rangle}\delta^4(p)=
\frac{\langle 12\rangle^2\langle 13\rangle}{\langle 23\rangle}\delta^4(p).
\end{equation}
In the second equality of (\ref{22oct23}) we used the momentum conservation in the form
\begin{equation}
\label{22oct24}
|2]\langle 2| = -|1] \langle 1| -|3] \langle 3|.
\end{equation}
The second term in (\ref{22oct22}) is treated similarly. Eventually, we end up with 
\begin{equation}
\label{22oct25}
{\cal A}_3\left(-\frac{3}{2},-\frac{1}{2},0\right) = (2\pi)^4 \frac{\langle 12\rangle^2\langle 13\rangle}{\langle 23\rangle}\delta^4(p).
\end{equation}
This example illustrates a typical computation of a three-point amplitude from the action. Higher-point amplitudes will not be discussed in the present paper.

\subsection{Amplitudes from symmetries}
\label{sec:2.3}

Instead of computing amplitudes from the action, one can study constraints imposed on them by symmetry considerations. It turns out that at the level of three-point amplitudes, for a fixed triplet of helicities, the amplitude is fixed by symmetries  up to an overall factor -- a coupling constant \cite{Benincasa:2007xk}. We will use analogous arguments in AdS, so let us briefly review this analysis in flat space. 

To start, translation invariance
\begin{equation}
\label{23octx1}
({\cal P}_{1|\alpha\dot\alpha}+{\cal P}_{2|\alpha\dot\alpha}+{\cal P}_{3|\alpha\dot\alpha})\;{\cal A} =0
\end{equation}
implies 
\begin{equation}
\label{23octx2}
{\cal A} = {\cal M}\;(2\pi )^4 \delta(P).
\end{equation}
Lorentz invariance requires that all spinor indices are contracted into spinor products, hence, 
\begin{equation}
\label{23octx3}
{\cal M} = {\cal M}(\langle ij\rangle, [ij]).
\end{equation}
Next, we further explore the structure of the right hand side of (\ref{23octx3}).

Momentum conservation together with the on-shell conditions for three-point amplitudes implies $p_i\cdot p_j=0$ for all pairs of particles, which in terms of spinors reads
\begin{equation}
\label{23oct1}
\langle 12\rangle [12]=0, \qquad  \langle 23\rangle [23]=0, \qquad \langle 31\rangle [31]=0. 
\end{equation}
Clearly, this entails that at least two spinor products of the same type are vanishing. Let us assume that 
\begin{equation}
\label{23oct2}
\langle 12\rangle =0, \qquad \langle 23\rangle =0.
\end{equation}
The spinor product $\langle ij \rangle$ vanishes iff $|i\rangle$ and $|j\rangle$ are parallel. Then (\ref{23oct2}) implies that $|1\rangle$ is parallel to $|2\rangle$ and $|2\rangle$ is parallel to $|3\rangle$. Together this entails that $|1\rangle$ is parallel to $|3\rangle$ and 
\begin{equation}
\label{23oct3}
\langle 13 \rangle =0.
\end{equation}
Hence, we conclude that the kinematics of massless three-point amplitudes implies that at least one type of spinor products vanishes simultaneously for all pairs of particles. For real momenta $[ij]=0$ implies $\langle ij\rangle=0$ and vice versa. Then all spinor products vanish and ${\cal M}$ can only be a constant. Taking into account (\ref{22oct3}), (\ref{22oct4}) one finds that this constant amplitude corresponds to three scalar fields, while all amplitudes for spinning fields are vanishing. 
To have non-trivial amplitudes for spinning fields, one allows momenta to be complex. This leaves one with an option of setting either all $\langle ij\rangle$ to zero and leaving $[ij]$ non-vanishing or vice versa. This implies that ${\cal M}$ splits into the holomorphic and antiholomorphic parts
\begin{equation}
\label{23oct4}
{\cal M}={\cal M}(\langle ij\rangle) + {\cal M}([ij]).
\end{equation}

Finally, let us fix helicities $h_i$ of particles on external lines.
 Helicity constraints (\ref{22oct3}), (\ref{22oct4}) on the fields carry over to the amplitude itself, so we obtain
\begin{equation}
\label{23oct5}
\frac{1}{2}\left(\Bar{\lambda}^i_{\dot{\alpha}}\frac{\partial}{\partial \Bar{\lambda}^i_{\dot{\alpha}}}-\lambda^i_{\alpha}\frac{\partial}{\partial \lambda^i_{\alpha}}\right){\cal M}(h_1,h_2,h_3)= h_i{\cal M}(h_1,h_2,h_3), \qquad \forall i.
\end{equation}
This implies
\begin{equation}
\label{23oct6}
{\cal M}(h_1,h_2,h_3) = g_{h} [12]^{d_{12,3}}[23]^{d_{23,1}}[31]^{d_{31,2}} + g_{a} \langle 12\rangle^{-d_{12,3}} \langle 23\rangle^{-d_{23,1}} \langle 31 \rangle^{-d_{31,2}},
\end{equation}
where $g_h$ and $g_a$ are two arbitrary coupling constants and
\begin{equation}
\label{23oct7}
d_{12,3}\equiv h_1+h_2-h_3, \qquad d_{23,1}\equiv h_2+h_3-h_1, \qquad d_{31,2} \equiv h_3+h_1-h_2.
\end{equation}
By demanding that the amplitude is non-singular in the limit of real momenta we find 
\begin{equation}
\label{23oct8}
\begin{split}
g_h=0 \quad\text{for} \quad h<0,\\
g_a=0 \quad\text{for} \quad h>0,
\end{split}
\end{equation}
where $h\equiv h_1+h_2+h_3$ is the total helicity.

To summarize,  Poincare covariance implies that three-point amplitudes of massless particles are given by (\ref{23octx2})
with ${\cal M}$ defined in (\ref{23oct6}). Moreover, when the total helicity is positive, only the anti-holomorphic part may be non-vanishing, while for negative total helicity, only the holomorphic part can be non-trivial. In particular, when the total helicity is zero, both terms in (\ref{23oct6}) are allowed.

It is interesting to compare (\ref{23oct8}) with the analogous conditions found in the light-cone deformation procedure \cite{Metsaev:1991mt}. In the light-cone approach these conditions result from imposing locality and differ in one respect: vertices with the total helicity being zero require non-local boost generators, unless all helicities are vanishing. In other words, unlike the spinor-helicity approach, in the light-cone deformation procedure vertices with the total helicity zero are not admissible, unless all fields are scalars\footnote{One can argue away these amplitudes in the spinor-helicity representation as well, but this requires to analyze consistency of higher-point functions \cite{Arkani-Hamed:2017jhn}.}.

\section{Massless Representations in AdS${}_4$}
\label{sec:3}

In the following sections we will generalize the flat space discussion reviewed in the previous section to AdS${}_4$ space. Our starting point is a deformation of (\ref{22oct1}) to AdS${}_4$ space given by
\begin{equation}
\label{23oct9}
\begin{split}
\mathcal{J}_{\alpha \beta}&=i \left(\lambda_{\alpha}\frac{\partial}{\partial\lambda^{\beta}}+\lambda_{\beta}\frac{\partial}{\partial\lambda^{\alpha}}\right),\\
\mathcal{\bar{J}}_{\dot{\alpha}\dot{\beta}}&=i\left(\bar{\lambda}_{\dot{\alpha}}\frac{\partial}{\partial\bar{\lambda}^{\dot{\beta}}}+\bar{\lambda}_{\dot{\beta}}\frac{\partial}{\partial\bar{\lambda}^{\dot{\alpha}}}\right),\\
\mathcal{P}_{\alpha\dot{\alpha}}&=\lambda_{\alpha}\bar{\lambda}_{\dot{\alpha}}-\frac{1}{R^2} \frac{\partial}{\partial\lambda^{\alpha}}\frac{\partial}{\partial\bar{\lambda}^{\dot{\alpha}}}.\\
\end{split}
\end{equation}
It is not hard to check that the generators given above form the familiar algebra of isometries of AdS${}_4$ with $R$ being the AdS radius. As in flat space, all generators commute with the helicity operator (\ref{22oct3}), which allows us to split the representation space into subspaces with definite helicity. Moreover, it is straightforward to check that $h=\pm s$ have the right values of the Casimir operators for a massless spin $s$ representation. In particular, the value of the quadratic Casimir operator
\begin{equation}
\label{23oct10}
    {\cal C}_2(so(3,2))\equiv \frac{1}{2}J_{AB}J^{AB}=\frac{R^2}{2}\epsilon^{\alpha\beta}\epsilon^{\dot{\alpha}\dot{\beta}}\mathcal{P}_{\alpha\dot{\alpha}}\mathcal{P}_{\beta\dot{\beta}}+\frac{1}{4}\epsilon^{\alpha\delta}\epsilon^{\beta\rho}\mathcal{J}_{\alpha\beta}\mathcal{J}_{\delta\rho}+\frac{1}{4}\epsilon^{\dot{\alpha}\dot{\delta}}\epsilon^{\dot{\beta}\dot{\rho}}\mathcal{\bar{J}}_{\dot{\alpha}\dot{\beta}}\mathcal{\bar{J}}_{\dot{\delta}\dot{\rho}},
\end{equation}
where $J_{AB}$ are the standard $so(3,2)$ generators,
for (\ref{23oct9}) is
\begin{equation}
\label{23oct11}
     {\cal C}_2(so(3,2))=2(h^2-1).
\end{equation}
This realization of massless representations in AdS${}_4$ is widely used in higher-spin theories and is often referred to as the twisted adjoint representation \cite{Bekaert:2005vh,Didenko:2014dwa}.

\section{AdS${}_4$ Geometry}
\label{geom}

Before moving to plane wave solutions, let us first choose convenient coordinates and introduce the necessary elements of the background geometry.

 For our purposes it will be helpful to make  Lorentz symmetry manifest. For this reason we choose coordinates, that may be regarded as a generalization of the stereographic coordinates on a sphere to AdS space. To be more precise, one can start from the familiar description of AdS space as a hyperboloid 
\begin{equation}
\label{23oct12}
X^M X_M = -R^2, \qquad M=0,1,\dots, 4
\end{equation}
embedded into a five-dimensional space with flat metric ${\rm diag}(-1,1,1,1,-1)$. By making a stereographic projection from $(0,0,0,0,-R)$ to $X^4=0$ hyperplane followed by a rescaling by a factor of two, we end up with new coordinates $x^a$, related to the ambient coordinates by
\begin{equation}
\label{23oct13}
    X^a=\frac{4x^aR^2}{4R^2-x^2}, \qquad X^4=R\frac{4R^2+x^2}{4R^2-x^2},
\end{equation}
\begin{equation}
\label{23oct14}
x^a = \frac{2 X^a}{1+X^4/R}.
\end{equation}
In these coordinates the metric reads
\begin{equation}
\label{23oct15}
ds^2=G^{-2}{\eta_{\mu\nu}dx^\mu dx^\nu},
\end{equation}
where
\begin{equation}
\label{24oct19}
G\equiv 1-\frac{x^2}{4R^2}
\end{equation}
and the AdS boundary is given by $x^2=4R^2$.

Stereographic projection (\ref{23oct13}), (\ref{23oct14}) maps $X^4<-R$ to $x^2>4R^2$, $-R<X^4<R$ to $x^2<0$ and $X^4>R$ to $0<x^2<4R^2$. We will refer to $x^2<4R^2$ and $x^2>4R^2$ as \emph{inner} and \emph{outer} patches respectively, while their union will be referred to as \emph{global} AdS.
 For points of AdS space with $X^4=-R$ the stereographic map degenerates: each generatrix of this cone maps to a point at infinity  along a null direction in intrinsic coordinates, while genuine infinite points in intrinsic coordinates correspond to $(0,0,0,0,-R)$ in ambient space. 

To deal with spinors in curved space one introduces a local Lorentz frame by means of the frame field
\begin{equation}
\label{23oct16x1}
g_{\mu\nu}=e_{\mu|}{}^{a} e_{\nu|}{}^b\eta_{ab},\qquad \eta_{ab} = e^{\mu|}{}_a e^{\nu|}{}_b g_{\mu\nu}, \qquad
e_{\mu|}{}^a e^{\nu|}{}_b=\delta^a_b.
\end{equation}
Here $\mu$ is the one-form index, while $a=0,1,2,3$ is a local Lorentz index.
We choose the local Lorentz basis to be
\begin{equation}
\label{23oct17}
e_{\mu|}{}^a=G^{-1} \delta^a_{\mu}.
\end{equation}
 As usual, the frame field is used to convert tensor fields from the coordinate to the local Lorentz basis and back, e.g.
\begin{equation}
\label{23oct18}
A^a=e^a_\mu A^\mu, \qquad A^\mu = e^\mu_aA^a.
\end{equation}
Moreover, local Lorentz indices are raised and lowered using the Minkowskian metric $\eta$.

We will use the following notation for the covariant derivatives of space-time tensors
\begin{equation}
\label{23oct19}
\nabla_\nu v_{\lambda}\equiv \partial_\nu v_{\lambda}-\Gamma_{\nu|}{}^\rho{}_{\lambda}v_\rho,\qquad
\nabla_\nu v^{\lambda}\equiv \partial_\nu v^\lambda +\Gamma_{\nu|}{}^\lambda{}_{\rho}v^\rho,
\end{equation}
where
\begin{equation}
\label{23oct20}
\Gamma_{\nu|}{}^{\rho}{}_{\lambda}=\frac{1}{2R^2}
G^{-1}\left(x_\nu \delta_\lambda^\rho +x_\lambda \delta_\nu^\rho - x^\rho \eta_{\nu\lambda} \right)
\end{equation}
are the Christoffel symbols for the torsion-free and metric-compatible connection. Covariant derivatives of Lorentz tensors are given by
\begin{equation}
\label{23oct21}
\nabla_\nu v^{a}\equiv \partial_\nu v^a +\omega_{\nu|}{}^{a,}{}_{b}v^b,
\qquad
\nabla_\nu v_{a}\equiv \partial_\nu v_{a}-\omega_{\nu|}{}^{b,}{}_{a}v_b=
\partial_\nu v_{a}+\omega_{\nu|}{}_{a,}{}^{b}v_b,
\end{equation}
where $\omega$ is the spin connection. Metric compatibility requires that it is antisymmetric
\begin{equation}
\label{23oct22}
0= \nabla_\nu \eta_{ab}=\partial_\nu \eta_{ab}-\omega_{\nu|}{}^{c,}{}_{a}\eta_{cb}-\omega_{\nu|}{}^{c,}{}_{b}\eta_{ac}=
-(\omega_{\nu|b,a}+\omega_{\nu|a,b}).
\end{equation}
Moreover, one requires that the frame field is covariantly constant
\begin{equation}
\label{23oct23}
0=\nabla_\nu e_{\mu|}{}^a = \partial_\nu e_{\mu|}{}^a -\Gamma_{\nu|}{}^{\rho}{}_{\mu}e_{\rho|}{}^a+
\omega_{\nu|}{}^{a,}{}_b e_{\mu|}{}^b,
\end{equation}
or, in other words, covariant derivatives (\ref{23oct19}) and (\ref{23oct21}) are compatible. One can solve (\ref{23oct23}) for $\omega$, which leads to
\begin{equation}
\label{23oct24}
\omega_{c|a,d}\equiv e^{\mu|}{}_c\omega_{\mu|a,d}=\frac{1}{2R^2}(\eta_{ca}x_d-\eta_{cd}x_a).
\end{equation}

The action of $so(3,2)$ on space-time tensors is realized by properly normalized Lie derivatives along Killing vectors. For example, deformed translations act on scalars as
\begin{equation} \label{MKV}
{\cal P}_i\varphi=-i\left(1+\frac{x^2}{4R^2}\right)\delta^\nu_i\frac{\partial}{\partial x^\nu}\varphi+i\frac{x_i}{2R^2}x^\mu\frac{\partial}{\partial x^\mu}\varphi.
\end{equation}
Isometries also act on local Lorentz indices. This action can be derived by requiring consistency of the action of the isometries on the space-time indices and relations like (\ref{23oct18}) that connect the two bases. Alternatively, one can notice that our choice of the frame field (\ref{23oct17}) is not invariant under isometries unless diffeomorphisms are supplemented with the appropriate local Lorentz transformations. In particular, considering that the frame field is a one-form, deformed translations act on it as follows
\begin{equation}
\label{24oct1}
    \left({{\cal P}_i}e\right)^a_{\mu}=-G^{-1}\frac{i}{2R^2}\left({x_{\nu}\delta^a_i-\eta_{i\nu}x^a}\right).
\end{equation}
To make the action of deformed translation consistent with our choice of the frame field, we must supplement them with local Lorentz transformations so that
\begin{equation}
\label{24oct2}
({\cal P}_i+\delta{\cal P}_i^{\rm L}) e=0.
\end{equation}
This leads to
\begin{equation}
\label{24oct3}
 (\delta{\cal P}_i^{\rm L} v)^a = (\zeta_i)^a{}_b v^b,\qquad   (\zeta_i)^a_{\hspace{0.1cm}b}=\frac{i}{2R^2}(x_b\delta_i^a-\eta_{ib}x^a).
\end{equation}

Local Lorentz indices can be converted to local spinor  ones using the standard vector-spinor dictionary. In particular, all formulae of this section can be translated to spinor notations. These can be found in appendix \ref{app:b}.

\section{Plane Waves for Field Strengths}
\label{sec:4}
As in flat space, to connect representation (\ref{23oct9}) realized in terms of differential operators in $sl(2,\mathbb{C})$ spinor space with the space-time fields, we need to find plane wave solutions. These will serve as intertwining kernels between the spinor-helicity and space-time representations\footnote{Interpretation of plane waves as intertwining kernels was used 
in  \cite{Fronsdal:1974ew,Fronsdal:1975eq}
to derive plane wave solutions for massive scalar and massive spin-$\frac{1}{2}$  fields in AdS${}_4$.}. 
In this section we discuss  plane wave solutions for  field strengths.

Let us consider helicity $h\ge 0$ field. Then, to saturate the homogeneity degree in spinor variables according to (\ref{22oct3}), (\ref{22oct4}), we consider an ansatz
\begin{equation}
\label{24oct4}
    F_{\dot\gamma_1\dots \dot\gamma_{2h}}(x,\lambda,\bar{\lambda})=\bar{\lambda}_{\dot{\gamma}_1}...\bar{\lambda}_{\dot{\gamma}_{2h}}f(x,\lambda,\bar{\lambda}),
\end{equation}
where $f$ has helicity zero. Next, Lorentz invariance requires that all spinor indices are covariantly contracted. This means that $f$ can only depend on two scalars $a$ and $b$, see (\ref{22oct7}). Finally, we have to require that deformed translations act consistently in the space-time and the spinor-helicity representations 
\begin{equation} \label{24oct5}
    \begin{split}
        &\left(-i\left(1+\frac{x^2}{4R^2}\right)(\sigma^a)_{\alpha\dot{\alpha}}\delta^\mu_a\frac{\partial}{\partial x^\mu}+i(\sigma^a)_{\alpha\dot{\alpha}}\frac{x_a}{2R^2}x^\mu\frac{\partial}{\partial x^\mu}+(\delta P^{\rm L}_{\alpha\dot{\alpha}})\right)\bar{\lambda}_{\dot{\gamma}_1}...\bar{\lambda}_{\dot{\gamma}_{2h}}d(a,b)\\
        &\hspace{6cm}=\left(\lambda_{\alpha}\bar{\lambda}_{\dot{\alpha}}-\frac{1}{R^2} \frac{\partial}{\partial\lambda^{\alpha}}\frac{\partial}{\partial\bar{\lambda}^{\dot{\alpha}}}\right)\bar{\lambda}_{\dot{\gamma}_1}...\bar{\lambda}_{\dot{\gamma}_{2h}}d(a,b).
    \end{split}
\end{equation}

Equation (\ref{24oct5}) has many components and for each of them it should be satisfied. Its independent components can be systematically found by taking $\lambda_\alpha$, $(x\bar\lambda)_\alpha\equiv x_{\alpha\dot\alpha}\bar\lambda^{\dot\alpha}$ to be the basis for holomorphic spinors and $\bar\lambda_{\dot\alpha}$, $(x\lambda)_{\dot\alpha}\equiv x_{\alpha\dot\alpha}\lambda^\alpha$ for antiholomorphic ones. In practice, however, 
using 
\begin{equation}
\label{20nov1}
x_{\alpha\dot\alpha}=\frac{b}{2a}\lambda_\alpha\bar\lambda_{\dot\alpha}+\frac{1}{a}x_{\alpha\dot\beta}\bar\lambda^{\dot\beta}\lambda^\beta x_{\beta\dot\alpha},
\end{equation}
we encounter only three different structures
\begin{equation}
\label{24oct6}
x_{\alpha\dot{\mu}}\bar{\lambda}^{\dot{\mu}} \bar{\lambda}_{\dot{\alpha}}
\lambda^{\beta}x_{\beta\dot{\gamma}_1}  \bar\lambda_{\dot \gamma_2}\dots \bar\lambda_{\dot \gamma_{2h}}+\dots,
 \qquad \lambda_{\alpha}\bar{\lambda}_{\dot{\alpha}} \bar\lambda_{\dot \gamma_1}\dots \bar\lambda_{\dot \gamma_{2h}},
 \qquad x_{\alpha\dot{\alpha}}\bar\lambda_{\dot \gamma_1}\dots \bar\lambda_{\dot \gamma_{2h}},
\end{equation}
which are, clearly, linearly independent. In (\ref{24oct6}) the omitted terms for the first structure make the expression symmetric in $\dot\gamma_i$.

The equation associated with the first spinor structure reads
\begin{equation} \label{24oct7}
    \frac{\partial d}{\partial a}=-\frac{i}{2}d,
\end{equation}
for which the solution is
\begin{equation}
\label{24oct8}
    d(a,b)=g(b)e^{-\frac{i}{2}a}.
\end{equation}
With (\ref{24oct8}) taken into account, the equation for the second structure is satisfied identically. Finally, considering the last structure, we find
\begin{equation}
\label{24oct9}
    \left(1+\frac{b}{8R^2}\right)\frac{dg}{db}=\frac{1+h}{8R^2}g.
\end{equation}
Within the class of genuine functions, the solution to the above equation is
\begin{equation}
\label{4dec1}
g(t)=C_1\left(1+\frac{b}{8R^2}\right)^{1+h}.
\end{equation}

From the analysis of the following sections it will be clear that plane waves associated with the solution (\ref{4dec1}) are not sufficient to generate all amplitudes consistent with symmetries. To be able to reproduce the missing amplitudes, one should also consider distributional solutions to (\ref{24oct9}).
Namely,  this equation can also be solved as
\begin{equation}
\label{24oct10}
    g(t)=C_{11}\left(1+\frac{b}{8R^2}\right)^{1+h}_++C_{12}\left(1+\frac{b}{8R^2}\right)^{1+h}_-,
\end{equation}
where $x_+\equiv x\theta(x)$ and $x_-\equiv -x\theta(-x)$. Indeed, (\ref{24oct9}) can be brought to the form
\begin{equation}
\label{4dec2}
xf'(x)=\lambda f(x).
\end{equation}
In the class of distributions, for $\lambda$ being not a negative integer (\ref{4dec2}) has the general solution  \cite{GS}
\begin{equation}
\label{4dec3}
f(x)=C_1 x_+^\lambda+C_2 x_-^\lambda.
\end{equation}
Discontinuity of solutions at $x=0$ is related to the fact that the higher-derivative term of the differential equation vanishes at this point.

Collecting everything together, we have found the following two independent plane wave solutions so far
\begin{equation}
\label{24oct11}
\begin{split}
 F^{\rm r|i}_{\dot\alpha_1\dots \dot\alpha_{2h}}&= \bar\lambda_{\dot\alpha_1}\dots \bar\lambda_{\dot \alpha_{2h}}\left(1-\frac{x^2}{4R^2}\right)_+^{1+h}e^{ipx},\\
F^{\rm r|o}_{\dot\alpha_1\dots \dot\alpha_{2h}}&= \bar\lambda_{\dot\alpha_1}\dots \bar\lambda_{\dot \alpha_{2h}}\left(1-\frac{x^2}{4R^2}\right)_-^{1+h}e^{ipx}.
\end{split}    
\end{equation}
Here label ${\rm r}$ refers to the fact that these solutions are \emph{regular} in the space-time, while labels ${\rm i}$ or ${\rm o}$ refer to the support of the solutions --- that is to the inner or to the outer patches. Along with these solutions, one can consider\footnote{Solution $F^{\rm r|g}$ was found in \cite{Bolotin:1999fa} using different methods. It is also implicitly present in the twistor literature \cite{Hitchin:1980hp,Plyushchay:2003gv}.}
\begin{equation}
\label{4dec5}
\begin{split}
F^{\rm r|g}_{\dot\alpha_1\dots \dot\alpha_{2h}}&= \bar\lambda_{\dot\alpha_1}\dots \bar\lambda_{\dot \alpha_{2h}}\left(1-\frac{x^2}{4R^2}\right)^{1+h}e^{ipx},
\end{split}    
\end{equation}
which is supported on the global patch. It is worth stressing, however, that for fermionic fields the analytic continuation across the interface between the patches is ambiguous due to the presence of square roots. Both these continuations are equally consistent with the analysis of symmetries we performed above.

Besides (\ref{24oct4}), one can consider other ways to saturate homogeneity degrees in spinor variables required by  the helicity constraint. Another way that leads to a solution is
\begin{equation}
\label{24oct12}
    F_{\gamma_1\dots \gamma_{2h}}(x,\lambda,\bar{\lambda})= (x\bar{\lambda})_{\gamma_1}...(x\bar{\lambda})_{\gamma_{2h}}d(a,b).\end{equation}
    Again, we get three independent spinorial structures, which give us three scalar equations. First, considering the equation for the structure
    \begin{equation}
    \label{24oct12x1}
    (x\bar\lambda)_\alpha \bar\lambda_{\dot\alpha}\lambda_{\gamma_1}(x\bar\lambda)_{\gamma_2}\dots(x\bar\lambda)_{\gamma_{2h}}+\dots
    \end{equation}
    we find 
    \begin{equation}
    \label{24oct12x2}
  b  \frac{\partial d}{\partial a}= {4R^2}i d,
    \end{equation}
    which, in the class of genuine functions,  gives
    \begin{equation}
    \label{24oct12x3}
    d(a,b)=g(b) {\rm exp}\left(i\frac{4R^2}{b}a \right).
    \end{equation}
    With (\ref{24oct12x3}) taken into account, the equation for 
    \begin{equation}
    \label{24oct12x4}
    \lambda_\alpha\bar\lambda_{\dot
    \alpha}(x\bar\lambda)_{\gamma_1}\dots(x\bar\lambda)_{\gamma_{2h}}
    \end{equation}
    is trivially satisfied. Finally, the equation for
    \begin{equation}
    \label{24oct12x5}
    x_{\alpha\dot\alpha}(x\bar\lambda)_{\gamma_1}\dots(x\bar\lambda)_{\gamma_{2h}}
    \end{equation}
    leads to
    \begin{equation}
    \label{24oct12x6}
    \left(1+\frac{b}{8R^2} \right)\frac{dg}{db}=-\frac{h}{8R^2}g-\frac{2h+1}{b}g.
    \end{equation}
    Again, focusing on solutions given by genuine functions, we find
    \begin{equation}
    \label{4dec4}
    g(t)= C_2b^{-1-2h}\left(1+\frac{b}{8R^2}\right)^{1+h}.
    \end{equation}
    
    In the distributional sense, solution (\ref{24oct12x3}), (\ref{4dec4}) is valid everywhere away from singular points of the equations. These are $b=0$ and $b=-8R^2$, where higher derivative terms in (\ref{24oct12x2}), (\ref{24oct12x6}) vanish. As in the example we considered before, one may expect that this solution can be truncated to domains, separated by these singularities. However, the fact that the solution  is itself singular at $b=0$ further complicates the analysis. Indeed, to solve (\ref{24oct12x2}), (\ref{24oct12x6}) in the distributional sense, one has to carefully define the associated distributions by integrating them against test functions and properly regularizing them and then learn how derivatives act on them. We leave this for future research. For now, we will write the solution as
     \begin{equation}
\label{24oct12x7}
    g(t)=C_{21}\;b^{-h}\left(1+\frac{8R^2}{b}\right)^{1+h}_++C_{22}\;b^{-h}\left(1+\frac{8R^2}{b}\right)^{1+h}_-.
\end{equation}   
    
The solutions of the second type are then given by
\begin{equation}
\label{24oct13}
\begin{split}
F^{\rm s|i}_{\alpha_1\dots\alpha_{2h}}
 &= \frac{(x\bar\lambda)_{\alpha_1}\dots  (x\bar\lambda)_{\alpha_{2h}}}{(x^2)^{{h}}}
\left(1-\frac{4R^2}{x^2}\right)_+^{1+{h}}e^{ipx \frac{4R^2}{x^2}},\\
F^{\rm s|o}_{\alpha_1\dots\alpha_{2h}}
 &= \frac{(x\bar\lambda)_{\alpha_1}\dots  (x\bar\lambda)_{\alpha_{2h}}}{(x^2)^{{h}}}\left(1-\frac{4R^2}{x^2}\right)_-^{1+h}e^{ipx \frac{4R^2}{x^2}}.
\end{split}    
\end{equation}
Here the label ${\rm s}$ refers to the fact that the solutions are \emph{singular} at $x^2=0$. One can also consider their linear combination of the form
\begin{equation}
\label{4dec6}
\begin{split}
F^{\rm s|g}_{\alpha_1\dots\alpha_{2h}}
 &= \frac{(x\bar\lambda)_{\alpha_1}\dots  (x\bar\lambda)_{\alpha_{2h}}}{(x^2)^{{h}}}
\left(1-\frac{4R^2}{x^2}\right)^{1+{h}}e^{ipx \frac{4R^2}{x^2}}.
\end{split}    
\end{equation}
Again, due to the presence of square roots, analytic continuation across the interfaces is ambiguous for fermionic fields.

Note that the inversion
 \begin{equation}
 \label{23oct16}
x^\mu  \leftrightarrow   x^\mu\, \frac{4R^2}{x^2}
\end{equation}
maps singular and regular solutions to each other, at least, for $x^2<0$, see appendix \ref{app:b} for details. It is easy to see that  when translated to ambient space terms, the inversion acts as the reflection with respect to the origin.

Flat space limit of (\ref{24oct11}), (\ref{4dec5}), (\ref{24oct13}), (\ref{4dec6}) is straightforward. In particular, $F^{\rm r|i}$ and $F^{\rm r|g}$ reduce to the familiar flat space plane waves when $R\to \infty$.
Let us also note another relation with the flat plane wave solutions. Massless representations are known to be conformally invariant in four dimensions \cite{Mack:1969dg,Mickelsson:1972tp,Angelopoulos:1980wg}. So is their description in terms of fields strengths, while the description in terms of potentials breaks conformal invariance except for the spin one case.
Considering that anti-de Sitter space is conformally equivalent to the Minkowski space, one can anticipate that upon dressing the solutions for field strengths in flat space with the appropriate powers of the conformal factor one should produce the associated solutions in AdS. This is exactly what we observe in (\ref{4dec5}): $F^{\rm r|g}$ is given by the flat space solution times a certain power of the conformal factor, see (\ref{23oct15}).

\section{Plane Waves for Potentials}
\label{sec:5}

In the present section we will study  plane wave solutions for potentials associated with the field strengths found in the previous section. For simplicity, we will focus on solutions of the type $F^{\rm r|g}$. Other solutions can be found similarly.

To find the potentials, we will consider the AdS counterparts of (\ref{22oct10}) and (\ref{22oct17}), take $F$ to be equal to $F^{\rm r|g}$ and then solve these equations for $\varphi$. As in flat space, this procedure allows us to define the potentials up to the gauge freedom. This gauge freedom can be fixed in many different ways and our goal is to fix it in a way that mimics the flat space spinor-helicity gauge (\ref{22oct12}), (\ref{22oct19}) as closely as possible. 

As we reviewed in section \ref{sec:2.1}, the spinor-helicity gauge has three remarkable properties: the potentials in this gauge are traceless, divergence-free and transverse to a given null vector. One can easily suggest natural generalizations of each ofthese conditions to AdS space. However, a simple inspection shows, that the resulting generalized conditions in AdS space cannot be satisfied simultaneously.

Having tried various possibilities, we found that it is the most natural to keep the condition of transversality to a given null vector intact, that is 
\begin{equation}
\label{24oct16}
q^{b}\varphi_{ba_2\dots a_h}=0, \qquad q^{b}\varphi_{\alpha|ba_2\dots a_h}=0
\end{equation}
for bosonic and fermionic fields respectively. There is a couple of reasons to do that. 
The main one is that the gauge condition (\ref{24oct16}) can always be achieved and, moreover,  fixes the gauge completely\footnote{This is the case when the appropriate boundary conditions are imposed at infinity. Otherwise, there is a residual symmetry, which will be discussed below.}. Another reason is that the transversality of the polarization vector to an auxiliary null vector that can be chosen arbitrarily -- in particular, as a momentum of a particle, appearing on one of the other external lines -- is an inherent feature of the spinor-helicity formalism that allows to simplify computations of four- and higher-point amplitudes. It seems reasonable to keep this feature in AdS space as well. We will sometimes refer to (\ref{24oct16}) as the \emph{spinor-helicity gauge}.

In the remaining part of this section we solve for the $q$-transverse plane wave potentials with spin up to two. Spin-0 and spin-$\frac{1}{2}$ cases are trivial as the potentials coincide with the field strengths. Due to conformal invariance, the analysis of the spin-1 case is identical to that in flat space. The remaining spin-$\frac{3}{2}$ and spin-2 cases turn out to be non-trivial.

\subsection{Spin 1}

As a warmup exercise, let us consider the spin-1 case.
In AdS space the spin-1 gauge transformation is
\begin{equation}
\label{24oct17}
    \delta A_{\mu}=\nabla_{\mu}\xi =\partial_{\mu}\xi
\end{equation}
and the associate gauge-invariant field strength reads
\begin{equation}
\label{24oct18}
    F_{\mu\nu}=\nabla_{\mu}A_{\nu}-\nabla_{\nu}A_{\mu}=\partial_{\mu}A_{\nu}-\partial_{\nu}A_{\mu}.
\end{equation}
From (\ref{4dec5}) we can see that the AdS field strength $F_{ab}$ is equal to the flat one times $G^2$. Then, converting its local Lorentz indices to the space-time ones, we find that  $F_{\mu\nu}$ in AdS space is identically equal to the field strength in flat space. So, we can solve (\ref{24oct18}) as in flat space, that is
\begin{equation}
\label{24oct20}
    A_{\mu}=-\frac{i}{2}\frac{(\sigma_{\mu})^{\dot{\alpha}\alpha}\mu_{\alpha}\bar{\lambda}_{\dot{\alpha}}}{\mu^{\beta}\lambda_{\beta}}e^{ipx}, \qquad      A_{\alpha\dot{\alpha}}=i\left(1-\frac{x^2}{4R^2}\right)\frac{\mu_{\alpha}\bar{\lambda}_{\dot{\alpha}}}{\mu^{\beta}\lambda_{\beta}}e^{ipx}.
\end{equation}

\subsection{Spin $\frac{3}{2}$}

There are two spin-$\frac{3}{2}$ potentials that we will denote $\psi_{\nu|\alpha}$ and $\bar\psi_{\nu|\dot\alpha}$. Their gauge transformations are given by 
\begin{equation}
\label{24oct21}
\delta \psi_{\nu|\alpha}=\nabla_\nu \xi_{\alpha}\pm \frac{1}{2R} e_{\nu|\alpha}{}^{\dot\alpha}\bar\xi_{\dot\alpha},
\qquad
\delta \bar\psi_{\nu|\dot\alpha}=\nabla_\nu \bar\xi_{\dot\alpha}\pm \frac{1}{2R} e_{\nu|\dot\alpha}{}^{\alpha}\xi_{\alpha}.
\end{equation}
The associated field strengths are 
\begin{equation}
\label{24oct22}
\begin{split}
F_{\mu\nu|\alpha}&=\nabla_\mu \psi_{\nu|\alpha}\pm \frac{1}{2R} e_{\mu|\alpha}{}^{\dot\alpha}\bar\psi_{\nu|\dot\alpha}-(\nu\leftrightarrow\mu)\\
\bar F_{\mu\nu|\dot\alpha}&=\nabla_\mu \bar\psi_{\nu|\dot\alpha}\pm \frac{1}{2R} e_{\mu|\dot\alpha}{}^{\alpha}\bar\psi_{\nu|\alpha}-(\nu\leftrightarrow\mu).
\end{split}
\end{equation}
As usual, both (\ref{24oct21}) and (\ref{24oct22}) can be found by making the most general ansatze involving no more than one derivative and fixing coefficients by requiring gauge invariance of field strengths. The sign ambiguity can be absorbed by a redefinition of the frame field. In what follows we will consider (\ref{24oct21}), (\ref{24oct22}) with the plus sign.

\subsubsection{Fixing an ansatz}

In the spinor language the transversality condition (\ref{24oct16}) for  $\psi$ and $\bar\psi$ reads
\begin{equation}
\label{26oct1}
\mu^\beta \bar\mu^{\dot\beta}\psi_{\beta\dot\beta|\alpha}=0, \qquad
\mu^\beta \bar\mu^{\dot\beta}\bar\psi_{\beta\dot\beta|\dot\alpha}=0.
\end{equation}
We are going to look for helicity $-\frac{3}{2}$ solution, which constrains the homogeneity degrees of both $\psi$ and $\bar\psi$ in $\lambda$ and $\bar\lambda$ according to (\ref{22oct3}), (\ref{22oct4}). Moreover, neither the on-shell value of $F$ nor operations involved in (\ref{24oct22}) depend on $\mu$ and $\bar\mu$. This implies that $\psi$ and $\bar\psi$ may only have homogeneity degrees zero in both $\mu$ and $\bar\mu$. These considerations together fix the ansatz for the potentials to be
\begin{equation}
\begin{split}
\label{15apr2}
\psi_{\beta\dot\beta|\alpha} = k_1 \frac{\lambda_\alpha \lambda_\beta \bar\mu_{\dot\beta}}{[\mu\lambda]}
+k_2 \frac{\lambda_{\alpha} \mu_{\beta}\bar\mu_{\dot\beta}}{[\mu \lambda] \langle\mu x\lambda]}+
k_3 \frac{\lambda_\alpha \mu_\beta \bar\lambda_{\dot\beta}}{\langle\mu x\lambda]^2}\langle \mu\lambda\rangle\\
+ k_4 \frac{\mu_\alpha \lambda_\beta \bar\mu_{\dot\beta}}{[\mu\lambda]\langle\mu x\lambda]}
+ k_5 \frac{\mu_\alpha \mu_\beta \bar\mu_{\dot\beta}}{[\mu\lambda] \langle\mu x\lambda]^2}
+ k_6 \frac{\mu_\alpha \mu_{\beta}\bar\lambda_{\dot\beta}}{\langle\mu x\lambda]^3}\langle \mu\lambda\rangle
\end{split}
\end{equation}
for $\psi$ and 
\begin{equation}
\begin{split}
\label{15apr4}
\bar\psi_{\dot\beta\beta|\dot\alpha} = l_1\frac{\bar\lambda_{\dot\alpha} \bar\lambda_{\dot\beta}\mu_{\beta}}{\langle\mu x\lambda]^3}\langle \mu\lambda\rangle^2
+ l_2 \frac{\bar\lambda_{\dot\alpha}\bar\mu_{\dot\beta}\mu_\beta}{[\mu \lambda]}\frac{\langle \mu \lambda\rangle}{\langle\mu x\lambda]^2} + l_3 \frac{\bar\lambda_{\dot\alpha}\bar\mu_{\dot\beta}\lambda_\beta}{[\mu \lambda]} \frac{\langle \mu\lambda\rangle}{\langle\mu x\lambda]}\\
+l_4 \frac{\bar\mu_{\dot\alpha}\bar\lambda_{\dot\beta}\mu_{\beta}}{[\mu \lambda]}\frac{\langle \mu\lambda\rangle}{\langle\mu x\lambda]^2}+
l_5 \frac{\bar\mu_{\dot\alpha}\bar\mu_{\dot\beta}\mu_\beta}{[\mu\lambda]^2 \langle\mu x\lambda]} +
l_6 \frac{\bar\mu_{\dot\alpha} \bar\mu_{\dot\beta}\lambda_\beta}{[\mu \lambda]^2}
\end{split}
\end{equation}
for $\bar\psi$. Here $k_i$ and $l_i$ are yet to be determined functions 
\begin{equation}
\label{26oct2}
k_i=k_i(a,b,c), \qquad l_i = l_i(a,b,c)
\end{equation}
of $a$ and $b$ defined in (\ref{22oct7}) and 
\begin{equation}
\label{30apr7}
c\equiv\frac{x_{\alpha\dot\alpha} \mu^\alpha \bar\mu^{\dot\alpha}}{\mu^\beta\lambda_\beta \bar\mu^{\dot\beta}\bar\lambda_{\dot\beta}}.
\end{equation}
All these variables have vanishing helicities and homogeneity degrees in $\mu$ and $\bar\mu$. It is not hard to show that all other scalar variables satisfying this property can be expressed in terms of $a$, $b$ and $c$.

\subsubsection{Solving for potentials}

The strategy of the derivation of $\psi$ and $\bar\psi$ is now straightforward. Namely, we take the ansatz (\ref{15apr2}), (\ref{15apr4}) for the potentials, evaluate the associated field strengths and equate them to a regular solution in global AdS space for $h=-\frac{3}{2}$, (\ref{4dec5}) . To be more precise, after translating field strengths (\ref{24oct22}) to the local Lorentz frame and converting the result to spinors, we should obtain
\begin{equation}
\begin{split}
\label{26oct1x1}
F_{\beta\gamma|\alpha}&=\lambda_\alpha\lambda_\beta\lambda_\gamma \left(1+\frac{b}{8R^2} \right)^{\frac{5}{2}} e^{-i \frac{a}{2}}, \qquad F_{\dot\beta\dot\gamma|\alpha}=0,\\
\bar F_{\beta\gamma|\dot\alpha}&=0, \qquad \qquad\qquad\qquad\qquad\qquad\quad \bar F_{\dot\beta\dot\gamma|\dot\alpha}=0.
\end{split}
\end{equation}

Considering that $F_{\beta\gamma|\alpha}$ is symmetric in $\beta$ and $\gamma$, this field strength has six independent components. 
 The same refers to $F_{\dot\beta\dot\gamma|\alpha}$, $F_{\dot\beta\dot\gamma|\dot\alpha}$ and $F_{\beta\gamma|\dot\alpha}$. Hence, in total the field strength has twenty-four components. As for the potentials, we use $\lambda$, $\bar\lambda$, $\mu$ and $\bar\mu$ as a basis for tensor structures. To obtain the contribution associated with each individual structure one can contract an expression with the appropriate combination of spinors, that annihilates all structures except a given one.
  For example, contraction with $\mu^\alpha\mu^\beta \mu^\gamma$ annihilates all components of $F_{\beta\gamma|\alpha}$ except the one, proportional to $\lambda_\alpha\lambda_\beta\lambda_\gamma$.

Proceeding along these lines, (\ref{26oct1}) gives twenty-four differential equations for twelve unknown functions $k_i$ and $l_i$ of three variables $a$, $b$ and $c$, which then should be solved. This computation is straightforward, but tedious. We give it in some detail in appendix \ref{app:c}, while here we only quote the result. Namely, we find that a particular solution for the potential is 
given by (\ref{15apr2}), (\ref{15apr4}) with
\begin{equation}
\label{1may1}
\begin{split}
k_1&= -i \sqrt{1+\frac{b}{8R^2}}\left(1+\frac{b}{8R^2} -\frac{1}{2R^2} ic \right)e^{-\frac{ia}{2}},\\
k_4 &= - \frac{1}{4R^2}(b-2ac)\sqrt{1+\frac{b}{8R^2}}e^{-\frac{ia}{2}},\\
l_6&= \frac{1}{R} \sqrt{1+\frac{b}{8R^2}}e^{-\frac{ia}{2}} 
\end{split}
\end{equation}
and other coefficient functions vanishing. The general solution of (\ref{26oct1x1}) is given by (\ref{1may1}) plus the general solution of the homogeneous equation, that is when the field strength is identically zero. Obviously, the latter solution corresponds to residual gauge transformations (\ref{24oct21}) for the gauge condition (\ref{26oct1}). As we demonstrate in appendix \ref{app:c}, these have different functional dependence on $a$, $b$ and $c$ compared to (\ref{1may1}), in particular, they do not allow  exponential dependence on $a$. Based on these considerations we make our choice of the particular solution (\ref{1may1}) of the inhomogeneous equation.

Finally, we note that the solution that we found, in fact, satisfies a stronger condition than (\ref{26oct1}), namely,
\begin{equation}
\label{26oct2x1}
 \bar\mu^{\dot\beta}\psi_{\beta\dot\beta|\alpha}=0, \qquad
 \bar\mu^{\dot\beta}\bar\psi_{\beta\dot\beta|\dot\alpha}=0.
\end{equation}
The same is also true in flat space. Given that (\ref{26oct2x1}) does not involve $\mu$, it makes sense to expect that the potentials  do not depend on $\mu$ at all, while they still may depend on $\bar\mu$.  Indeed, trading the holomorphic spinor basis $\{ \lambda_\alpha, \mu_\alpha\}$ for $\{\lambda_\alpha, (x\bar\lambda)_\alpha \}$, we find 
\begin{equation}
\begin{split}
\label{1may2}
\psi_{\beta\dot\beta|\alpha} &= -\left( i \left(1+\frac{b}{8 R^2} \right)^{\frac{3}{2}} +\frac{1}{4R^2}\frac{b}{a}\left(1+\frac{b}{8 R^2} \right)^{\frac{1}{2}} \right)
e^{-i\frac{a}{2}}  \frac{\lambda_\alpha \lambda_\beta\bar\mu_{\dot\beta}}{[\mu\lambda]}\\
&\qquad\qquad\qquad\qquad
 + \frac{1}{2R^2}\frac{1}{a} \left(1+\frac{b}{8 R^2} \right)^{\frac{1}{2}}
e^{-i\frac{a}{2}} \frac{x_{\alpha\dot\alpha}\bar\lambda^{\dot\alpha}\lambda_\beta \bar\mu_{\dot\beta}}{[\mu \lambda]^2 }\langle\mu x\lambda],
 \\
 \bar\psi_{\dot\beta\beta|\dot\alpha}&=\frac{1}{R}
 \left(1+\frac{b}{8 R^2} \right)^{\frac{1}{2}}
 e^{-i\frac{a}{2}}
 \frac{\bar\mu_{\dot\alpha}\bar\mu_{\dot\beta}\lambda_\beta}{[\mu\lambda]^2}.
\end{split}
\end{equation}

\subsection{Spin 2}

To find the helicity 2 potential we proceed in a similar manner. The gauge transformations are given by
\begin{equation}
\label{28oct1}
    \delta h_{\mu\nu}=\nabla_{\mu}\xi_{\nu}+\nabla_{\nu}\xi_{\mu}
\end{equation}
and the gauge-invariant field strength is
\begin{equation}
\label{28oct2}
\begin{split}
    F_{\mu\nu\rho\lambda}=&\nabla_{\rho}\nabla_{\mu}h_{\nu\lambda}-\nabla_{\rho}\nabla_{\nu}h_{\mu\lambda}-\nabla_{\lambda}\nabla_{\mu}h_{\nu\rho}+\nabla_{\lambda}\nabla_{\nu}h_{\mu\rho}\\
    &-\frac{1}{R^2}\left( g_{\nu\lambda}h_{\mu\rho}- g_{\nu\rho}h_{\mu\lambda}- g_{\mu\lambda}h_{\nu\rho}+ g_{\mu\rho}h_{\nu\lambda}\right).
\end{split}
\end{equation}
Next, we make the most general ansatz for the potentials in the spinor form that satisfies
\begin{equation}
\label{28oct3}
\mu^\beta \bar\mu^{\dot\beta}h_{\alpha\dot\alpha\beta\dot\beta}=0
\end{equation}
and has the appropriate homogeneity degrees in $\lambda$, $\bar\lambda$, $\mu$ and $\bar\mu$. This gives
\begin{equation}
\label{9nov7}
    \begin{split}
        h_{\alpha\beta,\dot{\alpha}\dot{\beta}}=&k_1\frac{\mu_{\alpha}\mu_{\beta}\Bar{\lambda}_{\dot{\alpha}}\Bar{\lambda}_{\dot{\beta}}}{\langle
        \mu\lambda\rangle^2}+k_2\frac{\mu_{\alpha}\mu_{\beta}(\Bar{\mu}_{\dot{\alpha}}\Bar{\lambda}_{\dot{\beta}}+\Bar{\mu}_{\dot{\beta}}\Bar{\lambda}_{\dot{\alpha}})}{\langle
        \mu\lambda\rangle^2 \langle\lambda x {\mu}]}+k_3\frac{\mu_{\alpha}\mu_{\beta}\Bar{\mu}_{\dot{\alpha}}\Bar{\mu}_{\dot{\beta}}}{\langle
        \mu\lambda\rangle^2 \langle\lambda x {\mu}]^2}\\
        &+k_4\frac{(\mu_{\alpha}\Bar{\mu}_{\dot{\beta}}\Bar{\lambda}_{\dot{\alpha}}\lambda_{\beta}+\Bar{\mu}_{\dot{\alpha}}\mu_{\beta}\lambda_{\alpha}\Bar{\lambda}_{\dot{\beta}})}{\langle
        \mu\lambda\rangle \langle\lambda x {\mu}]^2}[{\mu}{\lambda}]+k_5\frac{\Bar{\mu}_{\dot{\alpha}}\Bar{\mu}_{\dot{\beta}}(\mu_{\alpha}\lambda_{\beta}+\mu_{\beta}\lambda_{\alpha})}{\langle
        \mu\lambda\rangle \langle\lambda x {\mu}]^3}[{\mu}{\lambda}]\\
        &+k_6\frac{\Bar{\mu}_{\dot{\alpha}}\Bar{\mu}_{\dot{\beta}}\lambda_{\alpha}\lambda_{\beta}}{\langle\lambda x {\mu}]^4}[{\mu}{\lambda}]^2
    \end{split}
\end{equation}
with $k_i$ being arbitrary functions of $a$, $b$ and $c$.

 Then we evaluate the field strength (\ref{28oct2}) for this ansatz and require that all its components are vanishing except
\begin{equation}
\label{28oct4}
  \bar F_{\dot{\alpha}\dot{\beta}\dot{\gamma}\dot{\delta}}=\bar{\lambda}_{\dot{\alpha}}\bar{\lambda}_{\dot{\beta}}\bar{\lambda}_{\dot{\gamma}}\bar{\lambda}_{\dot{\delta}}\left(1-\frac{x^2}{4R^2}\right)^3e^{ip x}.
\end{equation}
The solution is then defined up to a residual gauge freedom that we fix in the same way as for the spin-$\frac{3}{2}$ case. Finally, as in the spin-$\frac{3}{2}$ case we find that the solution, actually, satisfies
\begin{equation}
\label{28oct5}
\mu^\beta h_{\alpha\dot\alpha\beta\dot\beta}=0
\end{equation}
and that the $\bar\mu$ dependence can be entirely eliminated. The end result is
\begin{equation}
\label{28oct6}
    \begin{split}
        h_{\alpha\dot{\alpha},\beta\dot{\beta}}=&-\left(1+\frac{b}{8R^2}-\frac{i}{2R^2}\frac{b}{a}\right)
        e^{-i\frac{a}{2}}
        \frac{\mu_{\alpha}\mu_{\beta}\bar{\lambda}_{\dot{\alpha}}\bar{\lambda}_{\dot{\beta}}}{\langle \mu \lambda \rangle^2}\\
        &-\frac{i}{2R^2}\frac{1}{a}e^{-i\frac{a}{2}}\frac{\mu_{\alpha}\mu_{\beta}(\lambda^{\gamma}x_{\gamma\dot{\alpha}}\bar{\lambda}_{\dot{\beta}}+\lambda^{\gamma}x_{\gamma\dot{\beta}}\bar{\lambda}_{\dot{\alpha}})}{\langle \mu \lambda \rangle^3}\langle \mu x {\lambda}]
    \end{split}
\end{equation}
Further details of this analysis are given in appendix \ref{app:d}.

\subsection{Higher-spin potentials}

A method of derivation of the plane-wave solutions in terms of  potentials we employed above  was straightforward, but tedious. Further generalizations of our results to the higher-spin case along the same lines are possible, but are expected to be even more complex, especially, due to the complicated form of the field strengths in AdS space, see e.g. \cite{Manvelyan:2007hv}. 
At the same time, simple form of plane wave solutions for the potentials that we obtained  in the lower-spin case suggests that there could be alternative and more economical approaches to a given problem.

For example, one may attempt to construct higher-spin potentials from lower-spin ones by applying helicity-changing operators. To be more precise, by making an ansatz for the most general operator with the right index structure and homogeneity degrees in $\lambda$, $\bar\lambda$, $\mu$ and $\bar\mu$ and requiring that it commutes with ${\cal P}_{\alpha\dot\alpha}$, one should be able to construct an operator that raises or lowers helicity of the potential by one. Then, by applying such operators multiple times to known potentials, one can generate a potential of any given helicity. One can further simplify this analysis by taking into account our observation that the dependence on one of the reference spinors drops out. The idea of helicity-changing operators will be successfully applied to generate three-point amplitudes in section \ref{sec:8}. Explicit analysis of the helicity-changing operators for the potentials will be given elsewhere. Finally, we mention that in a different gauge a somewhat implicit solution for the potentials associated with plane waves  $F^{\rm r|g}$ was given in \cite{Bolotin:1999fa}.

\section{Scattering Amplitudes from Space-Time Integrals}
\label{sec:6}
In anti-de Sitter space tree-level scattering amplitudes can be defined as the classical action evaluated on the solutions to the linearized equations of motion. The solutions of the linearized equations that we will be using in this computation are the plane waves that we derived in the previous section. This definition of amplitudes in AdS can be regarded as a straightforward generalization of the associated definition in flat space. It is also related to the holographic amplitudes computed by Witten diagrams by a mere change of a basis for the states appearing on external lines. In the following we will focus on amplitudes involving regular plane wave solutions.

All integrals that we will encounter will be of the following type \cite{GS}
\begin{equation}
\begin{split}
\label{28oct7}
{\cal I}^{\rm r|i}_\lambda& \equiv  \int d^4x{\left(1-\frac{x^2}{4R^2}\right)_+^{\lambda}} e^{i px}
\\
&\qquad \qquad =
  2^{\lambda+6}{\Gamma(\lambda+1)}\pi iR^4
\left[ e^{- i \pi(\lambda-\frac{1}{2})} \frac{K_{\lambda+2}(-2iR(p^2+i0)^{\frac{1}{2}})}{(-2iR(p^2+i0)^{\frac{1}{2}})^{\lambda+2}} 
- \text{c.c.} \right],\\
{\cal I}^{\rm r|o}_\lambda &\equiv  \int d^4x{\left(1-\frac{x^2}{4R^2}\right)_-^{\lambda} }e^{i px}\\
& \qquad\qquad = 2^{\lambda+6} {\Gamma(\lambda+1)}\pi iR^4
\left[ e^{i\frac{\pi}{2}} \frac{K_{\lambda+2}(-2iR(p^2+i0)^{\frac{1}{2}})}{(-2iR(p^2+i0)^{\frac{1}{2}})^{\lambda+2}} 
- \text{c.c.} \right],
\end{split}
\end{equation}
where c.c. denote complex conjugated terms. This formula is valid for all $\lambda$ except negative integers, for which the above integrals diverge. Somewhat formally, these integrals can be evaluated as 
\begin{equation}
\begin{split}
\label{28oct8}
{\cal I}^{\rm r|i}_{\lambda}=(2\pi)^4\left(1+\frac{\Box_p}{4R^2}\right)^{\lambda}_+\delta^4(p), \quad
{\cal I}^{\rm r|o}_{\lambda} =(2\pi)^4\left(1+\frac{\Box_p}{4R^2}\right)^{\lambda}_-\delta^4(p),
\end{split}
\end{equation}
which is the result of performing the Fourier transform according to a rule $x^2\to -\Box_P$. We will be primarily interested in the case of $\lambda$ being non-negative integer, $\lambda=n$. Then one can show that the formal computation
\begin{equation}
\label{28oct9}
{\cal I}^{\rm r|g}_{n}={\cal I}^{\rm r|i}_{n}+(-1)^n {\cal I}^{\rm r|o}_{n}= (2\pi)^4\left(1+\frac{\Box_p}{4R^2}\right)^{n}\delta^4(p),
\end{equation}
is  consistent with the rigorous formula (\ref{28oct7}), once the right hand sides of (\ref{28oct7}) are understood as distributions and appropriately regularized \cite{GS}.

\subsection{Simple examples}

In this section we evaluate a number of amplitudes, which, in effect, do not require the knowledge of the potentials in AdS space and for that reason can be computed easily.

\paragraph{Scalar self-interactions}
Consider a theory of a scalar field $\varphi$ with self-interaction
\begin{equation}
\label{29oct1}
S_n=\frac{1}{n!}\int d^4x \sqrt{-g} \varphi^n.
\end{equation}
We would like to compute the contact  $n$-point diagram.
Substituting regular global plane wave solutions (\ref{4dec5}) with helicity zero and computing the integral with the aid of (\ref{28oct7})-(\ref{28oct9}), we find
\begin{equation}
\label{29oct2}
{\cal A}_n^{\rm r|g}={\cal I}_{n-4}^{\rm r|g}.
\end{equation}
Similarly, for regular plane wave solutions supported on the inner and outer patches we obtain
\begin{equation}
\label{29oct3}
{\cal A}_n^{\rm r|i}={\cal I}_{n-4}^{\rm r|i}, \qquad {\cal A}_n^{\rm r|o}={\cal I}_{n-4}^{\rm r|o}.
\end{equation}

As it was noted before, singular solutions are related to regular ones by the inversion. Making the associated change of variables, one can compute amplitudes that involve only singular plane-wave solutions. In particular, one finds that
\begin{equation}
\label{11dec1}
{\cal A}_n^{\rm r|g}={\cal A}_n^{\rm s|g}.
\end{equation}
Similar relations hold for other patches. Amplitudes involving simultaneously regular and singular solutions are harder to compute.

\paragraph{Interactions involving field strengths}

Consider a theory with a vertex
\begin{equation}
\label{29oct4}
S_3=\frac{1}{2}\int d^4x \sqrt{-g}\varphi \bar F^{\dot\alpha\dot\beta}\bar F_{\dot\alpha\dot\beta}+{\rm c.c.}.
\end{equation}
Proceeding in a similar manner, for an antiholomorphic three-point amplitude  on different patches we find
\begin{equation}
\label{29oct5}
{\cal A}_3^{\rm r|i} = [ 23]^2  {\cal I}^{\rm r|i}_{1}, \quad
{\cal A}_3^{\rm r|o} = [ 23]^2  {\cal I}^{\rm r|o}_{1},\quad 
{\cal A}_3^{\rm r|g} = [ 23]^2  {\cal I}^{\rm r|g}_{1} 
\end{equation}
and similarly for the complex conjugate part. This example can be straightforwardly extended in two ways: to include higher-spin field strengths and to increase the number of fields in a vertex. Computation of amplitudes in all these examples is analogous.

\paragraph{Yang-Mills theory}

The Yang-Mills theory is classically conformally invariant, so one may expect that its amplitudes, at least at tree-level, are identical to those in flat space. Indeed, making the computation for lower-point cases, we find that all conformal factors cancel out and the AdS result coincides with the flat one. In particular, for the three-point amplitude we find
\begin{equation}
\label{29oct5x1}
{\cal A}_3^{\rm r|g}=\frac{[12]^3}{[23][31]}{\cal I}_{0}^{\rm r|g}
\end{equation}
and, similarly, for its complex conjugate.

\subsection{Genuine three-point amplitudes}

In the present section we will study more complicated examples. The interaction vertices we will consider cannot be written in terms of field strengths, nor are they conformally invariant. Our goal is to illustrate the genuine AdS spinor-helicity machinery, that is relevant for amplitudes that do not involve internal propagators. 

\subsubsection{Spin $0-\frac{1}{2}-\frac{3}{2}$ amplitude}
\label{sec:721}

Below we consider the AdS space version of the flat space computation presented in section \ref{sec:2.2}.
The AdS deformation of vertex (\ref{22oct20}) is given by
\begin{equation}
\label{29oct6}
S_3 = \int d^4 x \sqrt{-g} \left( \psi_{\mu|\alpha}\nabla^\mu \chi^\alpha \phi - \psi_{\mu|\alpha} \chi^\alpha \nabla^\mu \phi -
\frac{1}{2R} \bar\psi_{\mu|\dot\alpha} e^{\mu|\dot\alpha}{}_\alpha \chi^\alpha \phi \right) + {\rm c.c.}.
\end{equation} 
It is straightforward to check that it is invariant with respect to spin-$3/2$ gauge transformations (\ref{24oct21}), once the free equations of motion are taken into account
\begin{equation}
\label{29oct7}
\begin{split}
\left(\Box + \frac{2}{R^2} \right)\phi&\approx 0,\\
 e^{\mu|\alpha\dot\alpha} \nabla_\mu \chi_\alpha&\approx 0 \qquad \Rightarrow \qquad 
\left(\Box + \frac{3}{R^2} \right)\chi^\alpha\approx 0.
\end{split}
\end{equation}
Interaction vertex (\ref{29oct6}) is present in gauged supergravities \cite{deWit:2002vz}.

It is  easy to see that (\ref{29oct6}) cannot be expressed in terms of field strengths. Indeed, on-shell the only non-vanishing component of the field strength for $\psi$ carries three spinor indices that have nothing to be contracted with. One can also see that (\ref{29oct6}) is not conformally invariant simply by counting scaling dimensions. Hence, none of the simplifications encountered before take place in a given example and we have to deal with the full-fledged AdS spinor-heilicty machinery.

Let us proceed with the evaluation of the amplitude. First we plug into (\ref{29oct6}) explicit expressions for the metric, frame field and connections of the background geometry in our coordinates. This yields
\begin{equation}
\label{29oct8}
\begin{split}
S_3 &= \int d^4 x \left(1-\frac{x^2}{4 R^2} \right)^{-4} \left[ -\left(1-\frac{x^2}{4 R^2} \right) \psi^{\beta\dot\beta|\alpha}
\frac{\partial}{\partial x^{\beta\dot\beta}} \chi_{\alpha}\phi \right.\\
&\qquad\qquad+\left(1-\frac{x^2}{4 R^2} \right) \psi^{\beta\dot\beta|\alpha}
\chi_{\alpha}\frac{\partial}{\partial x^{\beta\dot\beta}} \phi 
-\frac{1}{8R^2} \psi^{\gamma\dot\gamma|}{}_\gamma x_{\beta\dot\gamma}\chi^\beta \phi
\\ 
&\qquad\qquad\qquad\qquad\qquad\qquad\left.-
\frac{1}{8R^2}\psi^{\gamma\dot\gamma|\alpha}x_{\alpha\dot\gamma}\chi_\gamma \phi
-\frac{1}{2R}\bar\psi^{\dot\alpha}{}_{\alpha|\dot\alpha}\chi^\alpha
\right]+\text{c.c.}.
\end{split}
\end{equation}
Next, we substitute  plane wave solutions (\ref{1may2}). A somewhat lengthy computation gives
\begin{equation}
\label{29oct9}
\begin{split}
{\cal A}_3^{\rm r|g} =& - \int d^4 x e^{ipx} \left(1-\frac{x^2}{4 R^2} \right)^{-1}[\mu 1]^{-2} \\
&\quad
\left[ \frac{i}{2}\left(i \left(1-\frac{x^2}{4R^2} \right) + \frac{x^2}{4R^2 p_1 x} \right)\left(1-\frac{x^2}{4 R^2} \right)
\langle 12\rangle [\mu 1](\langle 12\rangle [\mu 2]- \langle 13 \rangle [\mu 3])
\right.\\
&\qquad\qquad+\frac{i}{2}\frac{1}{4R^2 p_1 x}\left(1-\frac{x^2}{4 R^2} \right) \langle 2 x 1] \langle 1 x \mu] (\langle 12\rangle [\mu 2]- \langle 13 \rangle [\mu 3])\\
&\qquad\qquad\qquad+ \langle 12 \rangle \langle 1x\mu] [\mu 1]\left(-\frac{i}{4R^2}\left(1-\frac{x^2}{4 R^2} \right) - \frac{x^2}{32 R^4 p_1 x}  \right)\\
&\qquad\qquad\qquad\qquad\qquad\qquad\quad \left.
-\frac{1}{32 R^4 p_1 x} \langle 2 x 1] \langle 1 x\mu]^2 - \frac{1}{16 R^4} \langle 2 x \mu] \langle 1 x \mu]
 \right] .
\end{split}
\end{equation}
Now one can evaluate the $x$ integral by a formal replacement $x\to -i\partial_p$. The result can then be regarded as the AdS counterpart of (\ref{22oct22}).
However, to avoid derivatives in the denominator, we keep the integral in the form (\ref{29oct9}) and proceed with further simplifications. 

Our goal is to make manipulations analogous to (\ref{22oct23}) in flat space. To this end we need to  understand 
how the momentum conservation (\ref{22oct24}) translates to AdS space. 
 First, we note a trivial identity
\begin{equation}
\label{29oct10}
-\frac{2}{i} \frac{\partial}{\partial x^{\alpha\dot\alpha}} e^{ipx}=
(\lambda_{1\alpha}\bar\lambda_{1\dot\alpha}+\lambda_{2\alpha}\bar\lambda_{2\dot\alpha}+\lambda_{3\alpha}\bar\lambda_{3\dot\alpha}) e^{ipx}.
\end{equation}
Now, suppose, we would like to eliminate $|2\rangle |2]$ in favor of $|1\rangle |1]$ and $|3\rangle |3]$ as in flat space. In order to do that we just use 
\begin{equation}
\label{29oct11}
\lambda_{2\alpha}\bar\lambda_{2\dot\alpha} e^{ipx}
=(-\lambda_{1\alpha}\bar\lambda_{1\dot\alpha}-\lambda_{3\alpha}\bar\lambda_{3\dot\alpha} -\frac{2}{i} \frac{\partial}{\partial x^{\alpha\dot\alpha}}) e^{ipx}.
\end{equation}
Then the last term on the right hand side of (\ref{29oct11}) needs to be integrated by parts, thus differentiating the remaining part of the integrand. In flat space, due to translation invariance this contribution vanishes. Instead, in AdS space we have an explicit $x$-dependence, which results into additional non-trivial terms.

With this clarified, we proceed as in section \ref{sec:2.2}, except that we use the general formula (\ref{29oct11}) when the momentum conservation needs to be used. Namely, to eliminate $[\mu 2]$ in the numerators of the second and the third lines of
(\ref{29oct9}), we multiply the expression by $\langle23\rangle/\langle23\rangle$ and then integrate $|2\rangle |2]$ by parts using (\ref{29oct11}). Similar manipulations are then done with terms involving $[\mu 3]$. 

The remaining terms are simplified using the Schouten identities, see (\ref{19nov3}). In particular, in a given computation the following identities are used
\begin{equation}
\label{29oct12}
\begin{split}
\langle 2 x \mu] \langle 1x1] - \langle 2 x 1] \langle 1 x \mu] +\frac{1}{2} x_{\alpha\dot\alpha}x^{\alpha\dot\alpha} [\mu 1] \langle 1 2\rangle&=0,\\
\langle 12 \rangle \langle 3 x \mu] + \langle 1 x \mu] \langle 23\rangle - \langle 13 \rangle \langle 2 x \mu] &=0.
\end{split}
\end{equation}

Eventually, one finds
\begin{equation}
\label{29oct13}
{\cal A}_3^{\rm r|g}\left(-\frac{3}{2},-\frac{1}{2},0 \right) = \int d^4x \left(1-\frac{x^2}{4 R^2} \right) \frac{\langle 12 \rangle^2 \langle 31 \rangle}{\langle 23 \rangle} e^{ipx}
=\frac{\langle 12 \rangle^2 \langle 31 \rangle}{\langle 23 \rangle}  {\cal I}_1^{\rm r|g}.
\end{equation}
A similar result holds for the complex conjugate amplitude. As we will see in the next section, (\ref{29oct13}) is consistent with $so(3,2)$ covariance. This serves as a check of spin-$\frac{3}{2}$ potentials we found before.

\subsubsection{Spin $0-0-2$ amplitude}

Another example that we consider here is a cubic vertex that originates from the minimal coupling of a scalar field to gravity, see e.g. \cite{Bekaert:2010hk},
\begin{equation}
\label{29oct14}
    S_3=\int d^4x \sqrt{-g}h_{\mu\nu}j^{\mu\nu},
\end{equation}
where
\begin{equation}
\label{29oct15}
    j_{\mu\nu}=2(\nabla_{\mu}\phi)(\nabla_{\nu}\phi)-2\phi(\nabla_{\mu}\nabla_{\nu}\phi)-\frac{3}{R^2}g_{\mu\nu}\phi^2,
\end{equation}
$h_{\mu\nu}$ denotes the fluctuation of gravitational field around the AdS background $g_{\mu\nu}$ and $\phi$ is  a scalar field.
This vertex is not conformally invariant, neither it can be written in terms of field strengths. Hence, we have to deal with all the technicalities of a genuine spinor-helicity computation.

The analysis proceeds along the same lines as in the previous section. By substituting the plane wave solutions and explicit expressions for the background geometry, we find 
\begin{equation}
\label{29oct16}
    \begin{split}
        {\cal A}_3^{\rm r|g}&=\frac{1}{4}\int d^4x e^{ipx}\\
        &\left[\left(\frac{i}{2R^2}\frac{x^2}{p_3 x}+\frac{x^2}{4R^2}-1\right)\frac{1}{\langle \mu 3 \rangle^2}\left(\langle \mu 1 \rangle^2 [31]^2+\langle \mu 2 \rangle^2 [32]^2-2\langle\mu 1\rangle\langle\mu 2\rangle[31][32]\right)\right.\\
        &\quad+\frac{i}{R^2}\left(\frac{i}{2R^2}\frac{x^2}{p_3 x}+\frac{x^2}{4R^2}-1\right)\left(1-\frac{x^2}{4R^2}\right)^{-1}\frac{\langle \mu x 3] \langle \mu 1 \rangle [31]+\langle \mu x 3] \langle \mu 2 \rangle [32]}{\langle \mu 3 \rangle^2}\\
        &\quad-\frac{1}{2R^4}\left(\frac{i}{2R^2}\frac{x^2}{p_3 x}+\frac{x^2}{4R^2}-1\right)\left(1-\frac{x^2}{4R^2}\right)^{-2}\frac{\langle \mu x 3]^2}{\langle \mu 3 \rangle^2}\\
        &\quad+\frac{i}{2R^2}\frac{\langle \mu x 3]}{p_3 x\langle \mu 3\rangle^3}\left(\langle\mu 1\rangle^2\langle 3 x 1][31]+\langle\mu 2\rangle^2\langle 3 x 2][32]\right.\\
       &\qquad\qquad\qquad\qquad\qquad\qquad\qquad -\left.\langle\mu 1\rangle\langle\mu 2\rangle\langle 3 x 1][32]-\langle\mu 1\rangle\langle\mu 2\rangle\langle 3 x 2][31]\right)\\
        &\quad-\frac{1}{4R^4}\left(1-\frac{x^2}{4R^2}\right)^{-1}\frac{\langle \mu x 3]}{p_3 x \langle \mu 3\rangle^3}\left(x^2\langle 3 \mu\rangle\langle\mu 1\rangle[31]\right.\\
        &\qquad\qquad\qquad\qquad\qquad \left.+\langle\mu 1\rangle\langle\mu x 3]\langle 3 x 1]+x^2\langle 3 \mu\rangle\langle\mu 2\rangle[32]+\langle\mu 2\rangle\langle\mu x 3]\langle 3 x 2]\right)\\
        &\qquad\qquad\qquad\qquad\qquad\qquad \qquad\qquad\qquad+\left. \frac{i}{4R^6}\left(1-\frac{x^2}{4R^2}\right)^{-2}\frac{\langle \mu x 3]^2}{p_3x\langle \mu 3\rangle^2}x^2\right].
    \end{split}
\end{equation}
Next, using integration by parts, we trade $\langle \mu 1\rangle$ and $\langle \mu 2\rangle$ in the numerators for $\langle \mu 3\rangle$ as in section \ref{sec:721}. Along the way we used the Schouten identities
\begin{equation}
\label{4dec9}
\begin{split}
2  p_3 x  \langle \mu x 1]+{x^2}\langle \mu 3 \rangle [31]+{\langle \mu x 3]\langle 3 x 1]}&=0,\\
2   p_3 x \langle \mu x 2]-{x^2}\langle \mu 3 \rangle [23]+{\langle \mu x 3]\langle 3 x 2]}&=0,\\
    \langle 3 x 3][12]+\langle 3 x 1][23]+\langle 3 x 2][31]&=0,\\
    \langle \mu x 3][12]+\langle \mu x 1][23]+\langle \mu x 2][31]&=0.
    \end{split}
\end{equation}
Eventually, after a lengthy computation, we find
\begin{equation}
\label{29oct17}
    {\cal A}^{\rm r|g}_3(0,0,2)=-(2\pi)^4\frac{[23]^2[31]^2}{[12]^2}\left(1+\frac{\Box_P}{4R^2}\right)\delta^4(p)=
    -\frac{[23]^2[31]^2}{[12]^2}{\cal I}_1^{\rm r|g}.
\end{equation}
A similar result holds for the complex conjugate amplitude. As for the previous amplitude, (\ref{29oct17}) together with the analysis of the next section serves as a non-trivial consistency test of our formula for the spin-2 potential.

\begin{comment}
Having considered several examples, we can observe the following pattern. The spinor-helicity amplitudes in AdS turn out to have the form of a product of two factors: the first factor  consists of spinor products and saturates homogeneity degrees as required by the helicity constraints, while the second factor is ${\cal I}_{n}^{\rm r|i}$, ${\cal I}_{n}^{\rm r|o}$ or ${\cal I}_{n}^{\rm r|g}$, depending on the type of a vertex and the support of plane waves.

In particular, for three-point amplitudes with the total helicity $\sum h$ greater than zero we found two independent amplitudes
\begin{equation}
\label{29oct17x1}
{\cal A}_{\rm I}(h_1,h_2,h_3) = [12]^{d_{12,3}} [23]^{d_{23,1}}[31]^{d_{31,2}} {\cal I}_{\sum h-1}^{\rm r|i}
\end{equation}
and
\begin{equation}
\label{29oct17x2}
{\cal A}_{\rm II}(h_1,h_2,h_3) = [12]^{d_{12,3}} [23]^{d_{23,1}}[31]^{d_{31,2}} {\cal I}_{\sum h-1}^{\rm r|o}.
\end{equation}
Similarly, for $\sum h<0$ we found two amplitudes 
\begin{equation}
\label{29oct17x3}
{\cal A}_{\rm III}(h_1,h_2,h_3) = \langle 12\rangle^{-d_{12,3}} \langle 23\rangle^{-d_{23,1}}\langle 31\rangle^{-d_{31,2}} {\cal I}_{-\sum h-1}^{\rm r|i}
\end{equation}
and 
\begin{equation}
\label{29oct17x4}
{\cal A}_{\rm IV}(h_1,h_2,h_3) = \langle 12\rangle^{-d_{12,3}} \langle 23\rangle^{-d_{23,1}}\langle 31\rangle^{-d_{31,2}} {\cal I}_{-\sum h-1}^{\rm r|o}.
\end{equation}
\end{comment}

\section{Three-Point Amplitudes from Symmetries}
\label{sec:7}

In the previous section we computed several simple amplitudes in AdS${}_4$ using the spinor-helicity representation. In this section we will consider constraints imposed on three-point amplitudes purely from symmetry considerations. Our goal is to obtain a classification of three-point amplitudes analogous to that reviewed in section \ref{sec:2.3}.

To start, we remark that as in flat space, Lorentz invariance can be made manifest by contracting all spinor indices in a Lorentz-covariant manner. Next, we consider constraints imposed by requiring fixed helicities $h_i$ on external lines. These are  flat-space constraints (\ref{23oct5}). They can be solved as
\begin{equation}
\label{29oct18}
{\cal A}(h_1,h_2,h_3) =[12]^{d_{12,3}}[23]^{d_{23,1}}[31]^{d_{31,2}} f(x,y,z),
\end{equation}
where
\begin{equation}
\label{29oct19}
x\equiv [12]\langle 12\rangle, \quad y \equiv [23]\langle 23\rangle ,\quad z\equiv [31]\langle 31\rangle
\end{equation}
and $d$ were defined in (\ref{23oct7}).
What remains is to require invariance with respect to deformed translations, that is
\begin{equation}
\label{29oct20}
(\mathcal{P}_{1|\alpha\dot{\alpha}}+\mathcal{P}_{2|\alpha\dot{\alpha}}+\mathcal{P}_{3|\alpha\dot{\alpha}})\,{\cal A}(h_1,h_2,h_3)=0.
\end{equation}
This analysis is technically involved, so here we will just review the key steps, while further details can be found in appendix \ref{app:e}.

Invariance with respect to deformed translations (\ref{29oct20}) gives four second order differential equations for one unknown function $f$ of three variables, see (\ref{a11}). We were not able to find a systematic approach to solve them. Still, from direct computations of amplitudes in the previous section, one can anticipate that
\begin{equation}
\label{29oct17x1}
\begin{split}
{\cal A}_{\rm I}(h_1,h_2,h_3)& = [12]^{d_{12,3}} [23]^{d_{23,1}}[31]^{d_{31,2}} {\cal I}_{ h-1}^{\rm r|i},
\\
{\cal A}_{\rm II}(h_1,h_2,h_3) &= [12]^{d_{12,3}} [23]^{d_{23,1}}[31]^{d_{31,2}} {\cal I}_{ h-1}^{\rm r|o},
\\
{\cal A}_{\rm III}(h_1,h_2,h_3) &= \langle 12\rangle^{-d_{12,3}} \langle 23\rangle^{-d_{23,1}}\langle 31\rangle^{-d_{31,2}} {\cal I}_{- h-1}^{\rm r|i},
\\
{\cal A}_{\rm IV}(h_1,h_2,h_3)& = \langle 12\rangle^{-d_{12,3}} \langle 23\rangle^{-d_{23,1}}\langle 31\rangle^{-d_{31,2}} {\cal I}_{- h-1}^{\rm r|o}
\end{split}
\end{equation}
 gives four solutions to (\ref{29oct20}), which are, moreover, linearly independent. In this context, ${\cal I}$'s should be understood in the form (\ref{28oct7}) with $p^2 =-(x+y+z)$.
Next, we managed to show that  (\ref{29oct20}) do not have any other solutions in the class of genuine functions than those presented in (\ref{29oct17x1}).

To do that we consider an arbitrary non-singular point $(x_0,y_0,z_0)$ --- a point for which coefficients of higher-derivative terms in the equations do not vanish. Then, we consider equations (\ref{29oct20}) together with their derivatives and regard them as algebraic equations, expressing higher derivatives of $f$ at $(x_0,y_0,z_0)$ in terms of lower ones. The goal is to find how much of the initial data one has to specify at a given point, so that all derivatives of $f$ at this point and, hence, $f$ itself, are completely determined.
Proceeding in this manner, one can show that $f$ is uniquely specified by its value at $(x_0,y_0,z_0)$ and by values of its three derivatives. This implies that there are four linearly independent solutions to (\ref{29oct20}), which was to be demonstrated.

This argument is applicable once we are looking for solutions, given by genuine functions. At the same time, we may expect that (\ref{29oct20}) also has distributional solutions. Lorentz invariance imposes constraints on the domain on which these solutions are  supported. Considering also constraints from fixed helicities on external lines, an ansatz supported on $p=0$ reads
\begin{equation}
\label{30oct1}
{\cal A}(h_1,h_2,h_3) =[12]^{d_{12,3}}[23]^{d_{23,1}}[31]^{d_{31,2}} g(\Box_p)\delta(p).
\end{equation}
Imposing (\ref{29oct20}), we get a differential equation on $g$, which is a function of one variable.
This approach also leads to (\ref{29oct17x1}), where, ${\cal I}$'s appear  
in the form (\ref{28oct8}). Further details can be found in appendix \ref{app:e}.

To summarize, we find that in AdS${}_4$ once helicities are fixed, symmetry consideration alone leave room for only four consistent three-point amplitudes ${\cal A}_{\rm I}$, ${\cal A}_{\rm II}$, ${\cal A}_{\rm III}$ and ${\cal A}_{\rm IV}$ given in (\ref{29oct17x1}).
This result differs from the flat space classification discussed in section \ref{sec:2.3} in, essentially, one respect. Namely, the flat space momentum-conserving delta functions in AdS space get replaced with ${\cal I}$'s, for which we have two linearly independent possibilities (\ref{28oct8}) associated with two complementary patches of global AdS space. In addition, it is worth emphasizing that, though, we were able to generate amplitudes of the form ${\cal A}_{\rm I}$, ${\cal A}_{\rm II}$ from vertices only when the total helicity $h$ is positive and amplitudes of the form ${\cal A}_{\rm III}$, ${\cal A}_{\rm IV}$ when $h$ is negative, all four solutions are consistent with symmetry arguments independently of the value of $h$. This situation is reminiscent of that in flat space, where to reduce the number of solutions consitent with symmetries from two to one, we had to account for addition considerations, namely, require smooth limit for real momenta. Similarly, we can rule out ${\cal A}_{\rm I}$, ${\cal A}_{\rm II}$ for $ h <0$ and  ${\cal A}_{\rm III}$, ${\cal A}_{\rm IV}$ for $h>0$ by demanding regular flat limit. Note that these amplitudes are also singular, which can be seen from the gamma function factor in (\ref{28oct7}).

\section{Helicity-Changing Operators}
\label{sec:8}

Once a consistent amplitude is known, one can act on it with operators preserving background covariance, thus, generating other consistent amplitudes. This idea was used to establish relations between amplitudes in different theories in flat space, see e. g. \cite{Cheung:2017ems,Zhou:2018wvn,Bollmann:2018edb,Feng:2019tvb,Carrasco:2019qwr}.  
Similar phenomenon takes place for cubic vertices and three-point amplitudes of massless fields of any spin: all of them quite manifestly appear in the form of a seed scalar self-interaction acted upon by a sequence of differential operators, see e.g. \cite{Bengtsson:1986kh,Metsaev:1991mt,Benincasa:2007xk,Manvelyan:2010jr,Sagnotti:2010at}. The same idea can also be used to generate more complicated Witten diagrams and conformal correlators from simpler ones \cite{Paulos:2011ie,Costa:2011mg,Costa:2011dw,Karateev:2017jgd,Costa:2018mcg}. In the present section we will demonstrate how this approach can be implemented for three-point amplitudes in AdS${}_4$ in the spinor-helicity representation. We will call operators mapping one AdS${}_4$ spinor-helicity amplitude to another \emph{helicity-changing operators}.

The basic requirement for the helicity-changing operator is that it is Lorentz invariant. Taking into account that the $sl(2,\mathbb{C})$-invariant metric is antisymmetric, one quickly finds that one cannot construct non-trivial helicity-changing operators acting only on one external line. Focusing on the operators acting on two external lines, we construct
\begin{equation}
\label{6nov1}
\begin{split}
   D_{ij}^+&\equiv[ij]+\frac{1}{R^2}\epsilon_{\alpha\beta}\frac{\partial}{\partial\lambda_{i\alpha}}\frac{\partial}{\partial\lambda_{j\beta}},\\
      D_{ij}^-&\equiv\langle ij\rangle+\frac{1}{R^2}\epsilon_{\dot{\alpha}\dot{\beta}}\frac{\partial}{\partial \bar{\lambda}_{i\dot{\alpha}}}\frac{\partial}{\partial \bar{\lambda}_{j\dot{\beta}}}.
   \end{split}
\end{equation}
These operators are manifestly Lorentz-covariant. Moreover, the relative coefficients between the terms in $D$'s are chosen so that
\begin{equation}
\label{6nov2}
    [\mathcal{P}_{i|\alpha\dot{\alpha}}+\mathcal{P}_{j|\alpha\dot{\alpha}},D_{ij}^{\pm}]=0.
\end{equation}
Property (\ref{6nov2}) ensures that 
\begin{equation}
\label{6nov3}
(\dots+ {\cal P}_{i|\alpha\dot\alpha}+ \mathcal{P}_{j|\alpha\dot{\alpha}}+\dots ) {\cal A}=0 \qquad 
\Rightarrow \qquad 
(\dots+ {\cal P}_{i|\alpha\dot\alpha}+ \mathcal{P}_{j|\alpha\dot{\alpha}}+\dots ) D_{ij}^{\pm}{\cal A}=0.
\end{equation}
In other words, once a consistent amplitude ${\cal A}$ is available, $D$'s allow us to generate two other consistent amplitudes $D_{ij}^{\pm}{\cal A}$. Homogeneity degrees in spinors carried by $D_{ij}^\pm$ imply that $D_{ij}^+$ raises $h_i$ and $h_j$ by $\frac 12$, while $D_{ij}^{-}$ lowers them by $\frac{1}{2}$.
Operators $D^+_{ij}$ and $D^-_{ij}$ can be regarded as the AdS counterparts of flat space operators that multiply the amplitude with $[ij]$ and $\langle ij\rangle$ respectively.

Of course, one can apply helicity-changing operators in succession still producing consistent amplitudes. In particular, 
\begin{equation} 
\label{6nov4}
\begin{split}
    [D_{ij}^+,D_{ij}^-]&=\frac{2}{R^2}(H_i+H_j),\\
    [D_{ik}^+,D_{jk}^-]&=\frac{2}{R^2}H_{ij}, \qquad i\ne j
    \end{split}
\end{equation}
where $H_i$ was defined in (\ref{23oct5}) and 
\begin{equation}
\label{6nov5}
    2H_{ij}\equiv\Bar{\lambda}_{i\dot{\alpha}}\frac{\partial}{\partial \Bar{\lambda}_{j\dot{\alpha}}}-\lambda_{j\alpha}\frac{\partial}{\partial \lambda_{i\alpha}}.
\end{equation}
By construction, $H_{ij}$ is another helicity-changing operator. It raises $h_i$ by $\frac{1}{2}$ and lowers $h_j$ by $\frac{1}{2}$.

 A direct computation shows that
\begin{equation}
\label{6nov6}
\begin{split}
D^+_{12}{\cal A}_{\rm I}(h_1,h_2,h_3)&={\cal A}_{\rm I}\left(h_1+\frac{1}{2},h_2+\frac{1}{2},h_3\right),\\
D^-_{12}{\cal A}_{\rm I}(h_1,h_2,h_3)&=-\frac{1}{R^2}(h_1+h_2-h_3)(h_1+h_2+h_3-1){\cal A}_{\rm I}\left(h_1-\frac{1}{2},h_2-\frac{1}{2},h_3\right).
\end{split}
\end{equation}
Similar expressions can be found for ${\cal A}_{\rm II}$, ${\cal A}_{\rm III}$ and ${\cal A}_{\rm IV}$. In other words, we confirm that acting on a consistent three-point amplitude helicity-changing operators allows us to generate other consistent amplitudes.
This also serves as a consistency check for our derivation of three-point amplitudes using other methods. 

Though, we tested the idea of helicity-changing operators for three-point amplitudes only, property (\ref{6nov2}) guarantees that it should be valid for higher-point functions as well. It would be interesting to explore this further in future.

\section{Conclusion and Outlook}
\label{sec:9}

In a recent letter \cite{Nagaraj:2018nxq} we suggested a natural spinor-helicity formalism in AdS${}_4$ and made first steps in developing it. In particular, we suggested a convenient way of labelling the states appearing on the external lines of amplitudes by $sl(2,\mathbb{C})$ spinors. This labelling is defined by AdS${}_4$ plane wave solutions, which naturally extend the standard plane waves in flat space. With the plane wave solutions available, we computed several simple amplitudes. Next, we classified all consistent three-point amplitudes by requiring appropriate transformation properties with respect to the AdS${}_4$ isometry algebra $so(3,2)$. 

In the present paper we give technical details and proofs that were omitted in \cite{Nagaraj:2018nxq}. Moreover, previously, we only found plane wave solutions for  field strengths. In the present paper we proposed the AdS counterpart of the flat spinor-helicity representation for the potentials. The key property of this representation that we kept in AdS space is that the potentials are transversal to an auxiliary light-like vector, see (\ref{24oct16}). Once this gauge is fixed, we solved for the potentials associated with the plane wave solutions for field strengths we found before. We carried out the analysis for fields of spin up to two. Then we used these potentials to compute amplitudes for more non-trivial vertices. 
 These computations illustrate all technical aspects relevant for the computation of  diagrams without internal lines in our approach. Overall, this analysis is very reminiscent to the flat space one, except that in AdS space, due to the absence of translational invariance, the action explicitly depends on coordinates. This modifies the usual momentum conservation -- or, equivalently, integration by parts -- with extra terms, as well as requires to account for  additional Schouten identities, involving space-time coordinates.

The classification of three-point spinor-helicity amplitudes that we obtained is very similar to the flat-space one \cite{Benincasa:2007xk}. In particular, as expected, it contains additional amplitudes compared to those available in the approach that employs Lorentz tensors\footnote{For an incomplete list of references on cubic vertices in AdS, see \cite{Fradkin:1986qy,Vasilev:2011xf,Joung:2011ww,Boulanger:2012dx,Sleight:2016dba,Francia:2016weg,Karapetyan:2019psg}. Conformal three-point correlators of conserved currents were studied in \cite{Giombi:2011rz,Zhiboedov:2012bm}.}. This result is also consistent with the analysis in the light-cone gauge \cite{Metsaev:2018xip}. The resulting amplitudes are also very reminiscent of those in flat space: they only differ by what can be regarded as the AdS conformal factor raised to the power, that is defined by helicities on external lines. This suggests that the associated cubic vertices can be made conformally invariant by multiplying them with the appropriate power of the scalar field. It is natural to expect that chiral higher spin theories can be made conformally invariant in the same way. If this is true, one would obtain a new and simple way to relate higher-spin theories in flat and in AdS spaces: by promoting them to a parent conformal theory and then switching between the backgrounds by means of the Weyl transformations. Let us remind the reader, that the naive flat limit of higher spin theories in AdS is singular as the action contains negative powers of the cosmological constant. The approach that we sketched above may be free of this problem. 

Another related problem that would be interesting to explore is the following. In \cite{Sleight:2016dba} the complete cubic action for higher spin theories in AdS was defined from holography. For a vertex with a fixed triplet of spins, one can take the flat space limit smoothly if one first rescales it by the appropriate power of the cosmological constant \cite{Boulanger:2008tg,Bekaert:2010hw}. Then, in this limit only the highest derivative terms survive. The resulting highest derivative vertex in flat space can be compared to the cubic action derived independently in the light-cone gauge \cite{Metsaev:1991mt}. In \cite{Sleight:2016dba} it was found that the coupling constants of higher-derivative vertices in AdS${}_4$ and in four-dimensional flat space agree in the above sense\footnote{Based on the results for cubic couplings involving two scalar fields obtained in \cite{Bekaert:2015tva}, the conjecture that this matching should hold  in general was put forward in \cite{Skvortsov:2015pea}.}. The spinor-helicity formalism can be used to extend this analysis beyond the sector of higher-derivative vertices. Indeed, since spin corresponds to a pair of helicities, labeling of amplitudes with helicities is a more refined one than labelling with spins. In particular, for a fixed triplet of helicities we have a single consistent three-point amplitude, which is not the case for a triplet of spins. Using this observation, one can further split an AdS vertex with fixed spins into parts and rescale each part separately with the appropriate power of the cosmological constant, so that each part remains finite in the flat space limit. This would enable us to compare all the cubic action derived from holography and the flat space action in the light-cone gauge. In addition to the approach we suggested in the previous paragraph, this provides another way to relate higher spin theories in flat and AdS space backgrounds. 
More generally, it would be interesting to carry out the holographic reconstruction of higher-spin theories along the lines of \cite{Bekaert:2014cea,Bekaert:2015tva,Sleight:2016dba}, but in the spinor-helicity representation. This may be instructive to learn how the locality obstruction can be circumvented in flat space. 

An obvious direction to extend our results is to include higher-point functions. Already for four-point amplitudes at tree level we have two types of processes --- contact interactions and exchanges --- and it would be interesting to see how this difference manifests itself in the analytic structure of the associated amplitudes. Based on that one may then develop on-shell methods similar to those available in flat space.
It would also be interesting to extend other modern methods used for amplitudes' computations to AdS space. One step in this direction we have undertaken in the present paper: in section \ref{sec:8} we introduced helicity-changing operators, which are analogous to transmutation operators in flat space \cite{Cheung:2017ems}. In this context, it is also worth noting that three-point amplitudes for Yang-Mills and gravity in AdS satisfy a form of the double-copy relation \cite{Bern:2008qj,Bern:2019prr}: a combination of spinor products entering the gravity amplitude is just the square of the analogous factor for the Yang-Mills theory. Similar results were observed for other representations for AdS amplitudes and CFT correlators, see e.g. \cite{Farrow:2018yni}. 

The amplitudes we derived appear to be in the same representation as amplitudes computed using the twistor space techniques, see \cite{PR} for a general introduction to twistors and \cite{Adamo:2012nn,Skinner:2013xp,Adamo:2013tja,Adamo:2015ina,Haehnel:2016mlb,Adamo:2016ple,Adamo:2018srx} for computations of amplitudes of massless fields in  AdS${}_4$ space using this formalism. The reason is that the plane wave solutions we use for external lines are the same. The difference between our approaches is that we compute amplitudes from the usual space-time action, while in the twistor-space approach 
the amplitudes are computed from the twistor-space action. From this point of view, our approaches can be regarded as complementary. It is also worth noting that reformulation of the action in the twistor form is not always a simple task and such actions are not always available.

Another related approach was developed in \cite{Colombo:2012jx,Didenko:2012tv,Gelfond:2013xt}, where instead of $so(3,2)$ the whole higher spin symmetry was made manifest. The resulting amplitudes correspond to the scattering of the whole higher-spin multiplets. It would be interesting to decompose them into our basis and, in particular, identify  cubic couplings of higher-spin fields in this way.  Another closely related approach was recently discussed in \cite{Goncharov:2018exa}.

Finally, we comment on the relation of our approach to the usual holography. The main difference between the Witten diagrams and the amplitudes that we compute is that we use plane waves instead of bulk-to-boundary propagators for external lines of the diagrams. Both plane waves and bulk-to-boundary propagators provide a basis for solutions to free equations of motion. However, unlike bulk-to-boundary propagators, plane waves do not give a delta-function in a near-boundary limit, which means that the former should be identified not with local operators, but with operators smeared over the boundary\footnote{It is worth emphasizing another important difference: while we are dealing with Lorentzian signature, most of the literature on the AdS/CFT correspondence employs Euclidean signature.}. At the same time, our plane waves have an intuitive flat space limit, which makes this limit also straightforward at the level of amplitudes. In this regard, our plane wave solutions can be regarded as the scattering states \cite{Gary:2009ae,Heemskerk:2009pn,Fitzpatrick:2011jn,Maldacena:2015iua}, suitable for accessing flat-space physics from holography. They may also turn out to be convenient to deal with cosmological correlators upon the appropriate extension to de Sitter space.

\acknowledgments

We would like to thank E. Skvortsov for useful comments on the draft. The work of B. N. was supported by the Mitchell/Heep Chair in High Energy Physics. The work of D. P. was supported in part by  RSF Grant 18-12-00507.

\appendix
\section{Notations and Conventions}
\label{App:A}

The Pauli matrices are given by
\begin{equation}
\label{5nov3}
\sigma^0 = 
\left(\begin{array}{cccc}
1 &&& 0\\
0 && &1
\end{array}\right), \quad \sigma^1 = 
\left(\begin{array}{cccc}
0 & && 1\\
1& && 0
\end{array}\right), 
\quad 
 \sigma^2 = 
\left(\begin{array}{ccc}
0 && -i\\
i&& 0
\end{array}\right), \quad
 \sigma^3 = 
\left(\begin{array}{ccc}
1 && 0\\
0&& -1
\end{array}\right).
\end{equation}
These can be used to convert a vector index to a pair of spinor ones according to 
\begin{equation}
\label{5nov4}
p_{\alpha\dot\alpha}\equiv p_a (\sigma^a)_{\alpha\dot\alpha}.
\end{equation}
Here and throughout the paper we use Latin letters for Lorentz vector indices, while Greek letters from the beginning of the alphabet are used for spinor indices. Base indices are denoted by Greek letters from the middle of the alphabet. In flat space we use Cartesian coordinates, so we do not distinguish between local Lorentz and base indices.

For $p$ null, its spinor representation (\ref{5nov4}) factorizes
\begin{equation}
\label{5nov5}
p^ap_a=0 \qquad \Rightarrow \qquad p_{\alpha\dot\alpha}=\lambda_\alpha\bar\lambda_{\dot\alpha}.
\end{equation}
To raise and lower spinor indices we use the following convention
 \begin{equation}
\label{5nov6}
\lambda^\alpha = \epsilon^{\alpha\beta} \lambda_\beta, \qquad \lambda_\beta=\epsilon_{\beta\gamma} \lambda^\gamma,
\end{equation}
where
\begin{equation}
\label{5nov7}
\epsilon^{\alpha\beta}=\epsilon^{\dot\alpha\dot\beta}=
\left(
\begin{array}{cccc}
0&&& 1\\
-1&&&0
\end{array}
\right) = -\epsilon_{\alpha\beta}=-\epsilon_{\dot\alpha\dot\beta}.
\end{equation}
The same rule is used to raise and lower indices of the Pauli matrices.

 Relation (\ref{5nov4}) can be inverted to give
 \begin{equation}
 \label{5nov8}
 p_a=-\frac{1}{2}(\sigma_a)^{\dot\alpha\alpha}p_{\alpha\dot\alpha}.
 \end{equation}
 To this end one needs to use
 \begin{equation}
 \label{5nov9}
 (\sigma^a)_{\alpha\dot\alpha}(\sigma_a)_{\beta\dot\beta}=-2\epsilon_{\alpha\beta}\epsilon_{\dot\alpha\dot\beta},
 \qquad
  (\sigma^a)^{\alpha\dot\alpha}(\sigma_a)^{\beta\dot\beta}=-2\epsilon^{\alpha\beta}\epsilon^{\dot\alpha\dot\beta}.
 \end{equation}
 
 It is worth reminding the reader that antisymmetry of the $sl(2,\mathbb{C})$-invariant metric (\ref{5nov7}) leads to somewhat unusual properties of the spinor algebra. For example,
\begin{equation}
\label{5nov14}
\lambda_\alpha\psi^\alpha = -\lambda^\alpha\psi_\alpha.
\end{equation}
Clearly, this implies that the product of a spinor with itself vanishes. 

From the fact that the space where each spinor takes values is two-dimensional, it follows that antisymmtrization of a tensor with respect to a pair of indices is proprtional to the Levi-Civita tensor. The precise coefficient can be reconstructed by taking the trace of both parts. As a result, we get
\begin{equation}
\label{19nov2}
A_{\alpha\beta}-A_{\beta\alpha}=\epsilon_{\alpha\beta}A^\gamma{}_\gamma.
\end{equation}
Another consequence of the fact that the space of spinors is two-dimensional is the Schouten identity
\begin{equation}
\label{19nov3}
\lambda^\alpha \mu_\alpha \nu_\beta+\nu^\alpha \lambda_\alpha \mu_\beta+\mu^\alpha \nu_\alpha \lambda_\beta=0.
\end{equation}
It follows from (\ref{19nov2}) and the fact that antisymmetrization over three indices in two-dimensional space vanishes.

We define derivatives with respect to spinors in a natural way
\begin{equation}
\label{5nov15}
\frac{\partial \lambda^\alpha}{\partial \lambda^\beta}= \delta^\alpha_\beta, \qquad
\frac{\partial \lambda_{\alpha}}{\partial \lambda_{\beta}}= \delta_{\alpha}^{\beta}.
\end{equation}
 Then we find that
\begin{equation}
\label{5nov16}
\frac{\partial \lambda_\alpha}{\partial \lambda^\beta}=\epsilon_{\alpha\beta}, \qquad
\frac{\partial \lambda^\alpha}{\partial \lambda_\beta}=\epsilon^{\alpha\beta}.
\end{equation}
By comparing (\ref{5nov15}) with (\ref{5nov16}), we obtain
\begin{equation}
\label{5nov17}
\frac{\partial}{\partial \lambda^\alpha}=- \epsilon_{\alpha\beta}\frac{\partial}{\partial \lambda_\beta},
\end{equation}
which shows that indices of derivatives are  lowered with an extra minus sign compared to indices of spinors themselves (\ref{5nov6}). The same refers to raising indices and to antiholomorphic spinors.

 The vector-spinor dictionary (\ref{5nov4}), (\ref{5nov8}) can be extended to include tensors of any rank and symmetry. For example, for antisymmetric rank two tensor $C_{ab}=-C_{ba}$ one has 
 \begin{equation}
 \label{5nov10}
 C_{\alpha\dot\alpha,\beta\dot\beta}\equiv C^{ab}(\sigma_a)_{\alpha\dot\alpha}(\sigma_b)_{\beta\dot\beta}.
 \end{equation}
 One can then show that antisymmetry of $C^{ab}$ implies that $C_{\alpha\dot\alpha,\beta\dot\beta}$ is of the form
  \begin{equation}
 \label{5nov11}
 C_{\alpha\dot\alpha,\beta\dot\beta} = \epsilon_{\alpha\beta}\bar C_{\dot\alpha\dot\beta}+
 \epsilon_{\dot\alpha\dot\beta}C_{\alpha\beta},
 \end{equation}
 with $C_{\alpha\beta}$ and $\bar C_{\dot\alpha\dot\beta}$ symmetric.
  Here
 \begin{equation}
 \label{5nov12}
 \begin{split}
C_{\alpha\beta}=C_{\beta\alpha}\equiv \frac{1}{2}(\sigma_a)_\alpha{}^{\dot\gamma} (\sigma_b)_{\beta\dot\gamma}C^{ab},\\
\bar C_{\dot\alpha\dot\beta}=\bar C_{\dot\beta\dot\alpha} = \frac{1}{2}(\sigma_a)^\gamma{}_{\dot\alpha} (\sigma_b)_{\gamma\dot\beta}C^{ab}.
\end{split}
 \end{equation}
For real $C_{ab}$, $C_{\alpha\beta}$ and $\bar C_{\dot\alpha\dot\beta}$ are complex conjugate to each other. One can invert (\ref{5nov12}), which leads to
 \begin{equation}
\label{5nov13}
C_{ab} = \frac{1}{4} (\sigma_a)^{\dot\alpha\alpha}(\sigma_b)^{\dot\beta \beta}(\epsilon_{\alpha\beta}\bar C_{\dot\alpha\dot\beta}+
\epsilon_{\dot\alpha\dot\beta}C_{\alpha\beta}).
\end{equation}
Analogously, more general tensors can be treated, see e.g.  \cite{Didenko:2014dwa} for details.

Some other useful formulae in our conventions include
\begin{equation}
\label{5nov18}
\begin{split}
(\sigma_{a})_{\alpha\dot{\alpha}}({\sigma}^{b})^{\dot{\alpha}\alpha}=-2\delta^{b}_{a},\qquad
(\sigma_{a})_{\alpha\dot{\alpha}}({\sigma}^{a})^{\dot{\beta}\beta}=-2\delta^{\beta}_{\alpha}\delta^{\dot{\beta}}_{\dot{\alpha}},\\
({\sigma}^a)^{\dot\alpha\beta}(\sigma^b)_{\beta\dot\beta}+({\sigma}^b)^{\dot\alpha\beta}(\sigma^a)_{\beta\dot\beta}=-2\eta^{ab}\delta^{\dot{\alpha}}_{\dot{\beta}}.
\end{split}
\end{equation}

We will often use the standard shorthand notation
\begin{equation}
\label{5nov19}
\langle ij \rangle\equiv \lambda^i_\alpha\lambda^{j\alpha}=\lambda^i_\alpha\lambda^j_\beta\epsilon^{\alpha\beta}, \qquad
\left[ ij \right]\equiv \bar{\lambda}^i_{\dot{\alpha}}\bar{\lambda}^{j\dot{\alpha}}=\bar{\lambda}^i_{\dot{\alpha}}\bar{\lambda}^j_{\dot{\beta}}\epsilon^{\dot{\alpha}\dot{\beta}},
\end{equation}
where $i$ and $j$ label particles.
Moreover, we will also use notations of the following type
\begin{equation}
\label{5nov20}
\langle i x j] \equiv \lambda_i^\alpha x_{\alpha\dot\alpha} \bar\lambda^{\dot\alpha}_j,
\qquad \langle \lambda x \mu ] = \lambda^\alpha x_{\alpha\dot\alpha}\bar\mu^{\dot\alpha}.
\end{equation}

\section{AdS${}_4$ and Spinors}
\label{app:b}

In this appendix we give some of the formulae presented in section \ref{sec:3} in terms of spinors.

To start, we note that if 
\begin{equation}
\label{6nov7}
(\delta v)^a = \omega^{a,}{}_b v^b
\end{equation}
then
\begin{equation}
\label{6nov8}
(\delta v)_{\alpha\dot\alpha} =\frac{1}{2}\bar\omega_{\dot\alpha}{}^{\dot\beta}v_{\alpha\dot\beta}+\frac{1}{2}\omega_{\alpha}{}^\beta v_{\beta\dot\alpha}, \qquad 
(\delta v)^{\dot\alpha\alpha}=-\frac{1}{2}\bar\omega^{\dot\alpha}{}_{\dot\beta}v^{\dot\beta\alpha}-\frac{1}{2}\omega^{\alpha}{}_\beta v^{\dot\alpha\beta}.
\end{equation}
Here $\omega_{a,b}$ is an antisymmetric tensor and we use the standard vector-spinor dictionary reviewed in appendix \ref{App:A}. From (\ref{6nov8}) one can see that the action of an infinitesimal Lorentz transformation in spinor notations decomposes into two pieces each acting only on one type of spinor indices\footnote{Of course, there is an ambiguity in this decomposition. To be precise, one can add $i \varphi(x)\lambda_\alpha$ to $(\delta \lambda)_\alpha$  and $-i \varphi(x)\bar\lambda_{\dot\alpha}$ to $(\delta \bar\lambda)_{\dot\alpha}$ with $\varphi(x)$ real. This addition drops out from (\ref{6nov8}).}
\begin{equation}
\label{6nov9}
(\delta \lambda)_{\alpha} = \frac{1}{2}\omega_{\alpha}{}^\beta\lambda_{\beta}, \qquad (\delta\bar\lambda)_{\dot\alpha}=\frac{1}{2}\bar\omega_{\dot\alpha}{}^{\dot\beta}\bar\lambda_{\dot\beta}.
\end{equation}
By raising indices on both sides we can find how Lorentz transformations act on spinors with upper indices.

 Compatibility with vector formulae (\ref{23oct21}) then requires that covariant derivatives  act on spinor indices as follows
\begin{equation}
\label{6nov10}
\nabla_\mu \lambda_\alpha = \partial_\nu \lambda_\alpha + \frac{1}{2}\omega_{\mu|\alpha}{}^\beta\lambda_{\beta}, \qquad
\nabla_\mu \bar\lambda_{\dot\alpha} = \partial_\mu \bar\lambda_{\dot\alpha} +
\frac{1}{2}\bar\omega_{\mu|\dot\alpha}{}^{\dot\beta}\bar\lambda_{\dot\beta}.
\end{equation}
Similar formulae hold for spinors with upper indices. 

Now, let us find the spin connection in spinor notations. We start from (\ref{23oct24}) and convert the antisymmetric pair of indices to spinor ones using the standard dictionary. This gives
\begin{equation}
\label{6nov11}
\begin{split}
\omega_{c|\alpha\beta}&=\frac{1}{4R^2}\left( (\sigma_c)_{\alpha}{}^{\dot\gamma} x_{\beta\dot\gamma}-
x_{\alpha}{}^{\dot\gamma}(\sigma_c)_{\beta\dot\gamma}\right),
\\
\bar\omega_{c|\dot\alpha\dot\beta}&=\frac{1}{4R^2}\left( (\sigma_c)_{\dot\alpha}{}^{\gamma} x_{\gamma\dot\beta}-
x_{\dot\alpha}{}^{\gamma}(\sigma_c)_{\dot\beta\gamma}\right).
\end{split}
\end{equation}
In the following, we will find it convenient to convert the remaining Lorentz index to spinors too. This gives
\begin{equation}
\label{6nov12}
\begin{split}
\omega_{\gamma\dot\gamma|\alpha\beta}&=\frac{1}{2R^2}(\epsilon_{\gamma\alpha}x_{\beta\dot\gamma}+\epsilon_{\gamma\beta}x_{\alpha\dot\gamma}),
\\
\bar\omega_{\gamma\dot\gamma|\dot\alpha\dot\beta}&=\frac{1}{2R^2}\left(\epsilon_{\dot\gamma\dot\alpha}x_{\dot\beta\gamma}+\epsilon_{\dot\gamma\dot\beta}x_{\dot\alpha\gamma} \right).
\end{split}
\end{equation}

Finally, we present the spinor version of (\ref{24oct3})
\begin{equation}
\begin{split}
\label{1nov19}
    &(\delta_{\zeta} P_{\alpha\dot{\alpha}}\cdot\bar{\lambda})^{\dot{\beta}}=\frac{i}{4R^2}\left(\delta^{\dot{\beta}}_{\dot{\alpha}}x_{\alpha\dot{\gamma}}+\epsilon^{\dot{\beta}\dot{\delta}}x_{\alpha\dot{\delta}}\epsilon_{\dot{\gamma}\dot{\alpha}}\right)\bar{\lambda}^{\dot{\gamma}},\\
    &(\delta_{\zeta} P_{\alpha\dot{\alpha}}\cdot\lambda)^{\beta}=\frac{i}{4R^2}\left(\delta^{\beta}_{\alpha}x_{\gamma\dot{\alpha}}+\epsilon^{\beta\delta}x_{\delta\dot{\alpha}}\epsilon_{\gamma\alpha}\right)\lambda^{\gamma}.
\end{split}
\end{equation}

\paragraph{Inversion.} The change of coordinates
\begin{equation}
\label{25nov1}
x'^\mu = x^\mu \frac{4R^2}{x^2}
\end{equation}
induces the following action on tangent vectors 
\begin{equation}
\label{25nov2}
A'^{\mu'}(x')=\frac{\partial x'^{\mu'}}{\partial x^\mu}A^\mu (x).
\end{equation}
Explicit computation shows that 
\begin{equation}
\label{25nov3}
\frac{\partial x'^{\mu'}}{\partial x^\mu} =
\frac{4R^2}{x^2}\left(\delta^{\mu'}_\mu - 2\frac{x^{\mu'}x_\mu}{x^2} \right).
\end{equation}
By plugging (\ref{25nov3}) into (\ref{25nov2}) and going to the local Lorentz basis, we get
\begin{equation}
\label{25nov4}
A'^{a'}(x')= - \left(\delta^{a'}_a - 2\frac{x^{a'}x_a}{x^bx_b} \right)A^a(x).
\end{equation}
This relation can be converted to spinor notations to give 
\begin{equation}
\label{25nov5}
A'^{\alpha\dot\alpha}(x')=- \left( \delta^{\alpha}_\beta \delta^{\dot\alpha}_{\dot\beta}- \frac{x^{\alpha\dot\alpha}x_{\beta\dot\beta}}{x^bx_b}\right)A^{\beta\dot\beta}(x) =2\frac{x^\alpha{}_{\dot\beta}x^{\dot\alpha}{}_\beta}{x^{\gamma\dot\gamma}x_{\gamma\dot\gamma}} A^{\beta\dot\beta}(x).
\end{equation}
This action should be factorized into two pieces, each acting only on one type of spinor indices. This gives
\begin{equation}
\label{25nov8}
\lambda^\alpha= \sqrt{2}\frac{x^{\alpha}{}_{\dot\beta}}{\sqrt{x^{\gamma\dot\gamma}x_{\gamma\dot\gamma}}}\bar\lambda^{\dot\beta}, \qquad x^{\gamma\dot\gamma}x_{\gamma\dot\gamma}>0,
\end{equation}
and the action on holomorphic spinors is obtained by complex conjugation. It is not hard to see that for $x^{\gamma\dot\gamma}x_{\gamma\dot\gamma}<0$, (\ref{25nov5}) cannot be factorized into a product of transformations acting on individual spinors so that they remain complex conjugated to each other.

\section{Details on Spin $\frac{3}{2}$ Potential}
\label{app:c}

To start, we convert  field strengths (\ref{24oct22}) to the spinor notations. To be more precise, each pair of antisymmetric indices in $F_{\mu\nu|\alpha}$ and $\bar F_{\mu\nu|\dot\alpha}$ should be first transformed to the local Lorentz basis via (\ref{23oct18}) and then converted to spinors using the standard dictionary  (\ref{5nov12}). As a result, we get
\begin{equation}
\label{30apr1}
\begin{split}
F_{\beta\gamma|\alpha}
=\left(1-\frac{x^2}{4R^2} \right)\frac{\partial}{\partial x^{\beta\dot\epsilon}}\psi_{\gamma}{}^{\dot\epsilon}{}_{|\alpha} + \frac{2}{8R^2}
x_\gamma{}^{\dot\delta}\psi_{\beta\dot\delta|\alpha}\\
+
\frac{1}{8R^2}\left(\epsilon_{\beta\alpha}x^{\dot\delta\sigma}\psi_{\gamma\dot\delta|\sigma}-x^{\dot\delta}{}_\alpha
\psi_{\gamma\dot\delta|\beta} \right) \pm \frac{1}{2R}\epsilon_{\beta\alpha}\bar\psi_{\gamma\dot\delta|}{}^{\dot\delta}+(\beta\leftrightarrow\gamma),
\end{split}
\end{equation}
\begin{equation}
\label{30apr2}
\begin{split}
\bar F_{\dot\beta\dot\gamma|\alpha}
= \left(1-\frac{x^2}{4R^2} \right)\frac{\partial}{\partial x^{\tau\dot\beta}}\psi^{\tau}{}_{\dot\gamma|\alpha}
+\frac{2}{8R^2} x^\sigma{}_{\dot\beta} \psi_{\sigma\dot\gamma|\alpha}\\
+\frac{1}{8R^2}\left(x^\sigma{}_{\dot\beta} \psi_{\alpha\dot\gamma|\sigma}+x_{\alpha\dot\beta}\psi^\delta{}_{\dot\gamma|\delta} \right)\pm \frac{1}{2R}\bar\psi_{\alpha\dot\gamma|\dot\beta}+ (\dot\beta\leftrightarrow \dot\gamma).
\end{split}
\end{equation}
Making the complex conjugation of (\ref{30apr1}) and (\ref{30apr2}), we find the remaining components of the field strength. To achieve (\ref{30apr1}), (\ref{30apr2}) we needed to use explicit expressions for the frame field and the connection in spinor language, see appendix \ref{app:b}. Also note that derivatives are understood as follows
\begin{equation}
\label{7nov1}
\frac{\partial}{\partial x^a}f=(\sigma_a)^{\alpha\dot\alpha}\frac{\partial f}{\partial x^{\alpha\dot\alpha}}.
\end{equation}

Similarly, one finds a formula for gauge transformations of $\psi$
\begin{equation}
\label{30apr3}
\delta \psi_{\beta\dot\beta|\alpha}=-2\left(1-\frac{x^2}{4R^2} \right)\frac{\partial \xi_\alpha}{\partial x^{\beta\dot\beta}}-\frac{1}{4R^2} \epsilon_{\beta\alpha}x_{\beta\dot\gamma}\xi^\gamma-
\frac{1}{4R^2}\xi_\beta x_{\dot\beta\alpha}-\frac{1}{R}\epsilon_{\alpha\beta}\bar\xi_{\dot\beta}.
\end{equation}
Gauge variations of $\bar\psi$ can be obtained by complex conjugation.

As was explained in the main text, we then make an ansatz (\ref{15apr2}), (\ref{15apr4}) for the potential and evaluate field strengths. The computation turns out to be rather tedious, so we use computer algebra. Then we equate the field strength found to its on-shell value component by component. For example, an equation resulting from setting the $\lambda_\alpha\lambda_\beta\lambda_\gamma$ component to its on-shell value gives
\begin{equation}
\label{7nov2}
\lambda_\alpha\lambda_\beta\lambda_\gamma : \qquad 
\frac{c}{4R^2}k_1 - 4c
\left(1+\frac{b}{8R^2} \right)
\partial_b k_1 -2 
\left(1+\frac{b}{8R^2} \right)
\partial_a k_1=\left(1+\frac{b}{8R^2} \right)^{\frac{5}{2}}e^{-i\frac{a}{2}}.
\end{equation}
Similarly, for the $\mu_\alpha\lambda_\beta\lambda_\gamma$ component we find
\begin{equation}
\label{7nov3}
\mu_\alpha\lambda_\beta\lambda_\gamma : \qquad 
\frac{b-2ac}{4R^2} k_1-\frac{3c}{4R^2}k_4-\frac{1}{R}l_3 +4c
\left(1+\frac{b}{8R^2} \right)
 \partial_bk_4 +2
 \left(1+\frac{b}{8R^2} \right)
 \partial_ak_4=0.
\end{equation}
Equations associated with other 22 components are similar. 

To solve these equations, we first consider the homogeneous system, that is when the right hand sides in (\ref{26oct1x1}) are absent. In terms of component equations this implies that the right hand side of (\ref{7nov2}) should be set to zero, while all the remaining equations remain intact. Solutions of these equations, by construction, correspond to pure gauge modes. 

To find these pure gauge solutions, we make a general ansatz for gauge parameters, similar to the one we made for potentials
\begin{equation}
\label{7nov4}
\begin{split}
\xi_\alpha &= m_1 \frac{\lambda_\alpha\langle\mu\lambda\rangle}{\langle\mu x\lambda]}+m_2 \frac{\mu_\alpha \langle \mu \lambda\rangle}{\langle\mu x\lambda]^2},\\
\bar\xi_{\dot\alpha}&= n_1 \frac{\bar\lambda_{\dot\alpha}\langle \mu\lambda\rangle^2}{\langle\mu x\lambda]^2}
+n_2 \frac{\bar\mu_{\dot\alpha}\langle\mu\lambda\rangle}{\langle\mu x\lambda][\mu \lambda]}.
\end{split}
\end{equation}
Here $n_i$ and $m_i$ are arbitrary functions of $a$, $b$ and $c$.
By making a gauge variation
 (\ref{30apr3}) and requiring (\ref{26oct1}) --- which boils down to the vanishing of the components of $\psi$ along 
 $\lambda_\beta \bar\lambda_{\dot\beta}\lambda_\alpha$ and $\lambda_\beta \bar\lambda_{\dot\beta}\mu_\alpha$ ---
 we find to equations for  $n_i$ and $m_i$ of residual gauge transformations. By imposing (\ref{26oct1}) for $\bar\psi$, we get another two equations. Solving these equations, we find 
 \begin{equation}
 \label{7nov5}
 \begin{split}
 m_1(a,b,c)& =-\frac{m_1^r(c,b-2ac)}{\sqrt{b+8R^2}},\\
   m_2(a,b,c)&=-\frac{(b-2ac)m_1^r(c,b-2ac)+2c(b+8R^2)m_2^r(c,b-2ac)+4Rn_1^r(c,b-2ac)}{2c\sqrt{b+8R^2}}\\
   n_1(a,b,c)&=\frac{n_1^r(c,b-2ac)}{\sqrt{b+8R^2}},\\
   n_2(a,b,c,)&=\frac{2R m_1^r(c,b-2ac) -n_1^r(c,b-2ac)+c(b+8R^2)n_2^r(c,b-2ac)}{c\sqrt{b+8R^2}}.
  \end{split}
 \end{equation}
Here $m_i^r$ and $n_i^r$ are four arbitrary functions of two variables $c$ and $b-2ac$.

Having clarified how residual gauge transformations act, we proceed with solving equations for the potentials. 
Inhomogeneous equation (\ref{7nov2}) involves only $k_1$ and can be solved as follows
\begin{equation}
\label{7nov6}
k_1=-i\sqrt{1+\frac{b}{8R^2}}\left(1+\frac{b}{8R^2}-\frac{1}{2R^2}ic \right)e^{-i\frac{a}{2}}+
\sqrt{1+\frac{b}{8R^2}}r(c,b-2ac),
\end{equation}
with $r$ being an arbitrary function of $c$ and $b-2ac$. Clearly, the $r$-term in (\ref{7nov6}) gives a general solution of the homogeneous equation, that is when the right hand side of (\ref{7nov2}) is set to zero. These solutions should correspond to pure gauge potentials and we checked that, indeed, variation (\ref{30apr3}) with parameters given by (\ref{7nov4}), (\ref{7nov5}) contributes such a term. So, by further fixing the gauge symmetry, we can adjust $r$ in (\ref{7nov6}) in any convenient way.
We find it convenient to set $r$ to zero. One reason for that is that $r$ does not allow exponential ${\rm exp}(-ia/2)$ dependence
typical of plane waves, instead, featuring $a$ only in combination $b-2ac$.

Then we proceed with the remaining equations one after another. These can be solved systematically, as the system of equations admits ''lower-triangular form''.
To be more precise, some of them involve only one unknown function and can be immediately solved like (\ref{7nov2}). Each time we pick a particular solution so that terms without exponential behavior in $a$ are absent. Plugging these solutions into remaining equations we again find equations with only one unknown function and solve them. We proceed like that until all unknown functions are solved for. 
The end result is given in (\ref{1may1})

Finally, we note that the solution (\ref{15apr2}), (\ref{15apr4}), (\ref{1may1}) satisfies a stronger gauge condition (\ref{26oct2x1}), which suggests that it is, actually, $\mu$-independent. Using
\begin{equation}
\label{7nov7}
\mu_\alpha = \frac{\langle \mu \lambda\rangle}{\langle\lambda x \lambda]}x_{\alpha\dot\alpha}\bar\lambda^{\dot\alpha}+
\frac{\langle \mu x \lambda]}{\langle \lambda x \lambda]}\lambda_\alpha
\end{equation}
to eliminate $\mu$ in the $k_4$ and employing
\begin{equation}
\label{7nov8}
b-2ac = -2 \frac{\langle\lambda x \mu]\langle \mu x \lambda]}{\langle \mu \lambda\rangle [\mu \lambda]}
\end{equation}
we find that, indeed, $\mu$-dependence cancels. The final result is given in (\ref{1may2}).

\section{Details on Spin $2$ Potential}
\label{app:d}

Here we give some intermediate results of the computation of the spin-2 potential. 

First, we convert both the potential and the field strength to local Lorentz indices and then to spinor ones. For the spin-2 potential the standard dictionary reads
\begin{equation}
\label{9nov1}
    h_{ab}=\frac{1}{4}h_{\alpha\beta,\Dot{\alpha}\dot{\beta}}(\Bar{\sigma}_a)^{\dot{\alpha}\alpha}(\Bar{\sigma}_b)^{\dot{\beta}\beta}, \qquad 
    h_{\alpha\beta,\Dot{\alpha}\dot{\beta}}=(\sigma^a)_{\alpha\dot{\alpha}}(\sigma^b)_{\beta\dot{\beta}}h_{ab}.
\end{equation}
The field strengths has the symmetry of a window-shaped Young diagram. Namely, it satisfies 
\begin{equation} 
\label{9nov2}
    F_{ab,cd}=-F_{ab,dc}, \qquad 
    F_{ab,cd}=-F_{ba,cd}, \qquad 
    F_{ab,cd}=F_{cd,ab}
\end{equation}
and the Young symmetry condition. Converting each pair of antisymmetric indices to spinor ones using (\ref{5nov12}), we get
\begin{equation}
\label{9nov3}
\begin{split}
    \Bar{F}_{\dot{\alpha}\dot{\beta}\gamma\delta}=\frac{1}{4}(\sigma^a)^{\beta}_{\hspace{0.2cm}\dot{\alpha}}\hspace{0.1cm}(\sigma^b)_{\beta\dot{\beta}}\hspace{0.1cm}(\sigma^c)_{\gamma}^{\hspace{0.2cm}\dot{\delta}}\hspace{0.1cm}(\sigma^d)_{\delta\dot{\delta}}\hspace{0.1cm}F_{ab,cd}, 
\\
    F_{\dot{\alpha}\dot{\beta}\dot{\gamma}\dot{\delta}}=\frac{1}{4}(\sigma^a)^{\beta}_{\hspace{0.2cm}\dot{\alpha}}\hspace{0.1cm}(\sigma^b)_{\beta\dot{\beta}}\hspace{0.1cm}(\sigma^c)^{\delta}_{\hspace{0.2cm}\dot{\gamma}}\hspace{0.1cm}(\sigma^d)_{\delta\dot{\delta}}\hspace{0.1cm}F_{ab,cd}
    \end{split}
\end{equation}
and two other field strengths, that can be found by complex conjugation of (\ref{9nov3}). 

By using explicit expressions for the frame fields and spin connections in our coordinates, after a lengthy computation we find
\begin{equation}
\label{9nov4}
    \begin{split}
        F_{\dot{\alpha}\dot{\beta}\dot{\gamma}\dot{\delta}}=&\left(1-\frac{x^2}{4R^2}\right)^2\epsilon^{\alpha\beta}\epsilon^{\gamma\delta}
        \frac{\partial}{\partial x^{\dot{\alpha}\alpha}}\frac{\partial}{\partial x^{\dot{\gamma}\gamma}}h_{\beta\delta,\dot{\beta}\dot{\delta}}
        \\
        &-\frac{1}{4R^2}\left(1-\frac{x^2}{4R^2}\right)\epsilon_{\dot{\alpha}\dot{\delta}}\epsilon^{\beta\gamma}x^{\dot{\sigma}\sigma}\left[\frac{\partial}{\partial x^{\dot{\gamma}\gamma}}h_{\sigma\beta,\dot{\sigma}\dot{\beta}}+\frac{\partial}{\partial x^{\dot{\beta}\beta}}h_{\sigma\gamma,\dot{\sigma}\dot{\gamma}}\right]
       \\
        &+\frac{1}{4R^2}\left(1-\frac{x^2}{4R^2}\right)\left(-2+x^{\dot{\sigma}\sigma}\frac{\partial}{\partial x^{\dot{\sigma}\sigma}}\right)\epsilon_{\dot{\beta}\dot{\delta}}\epsilon^{\alpha\gamma}\hspace{0.1cm}h_{\alpha\gamma,\dot{\alpha}\dot{\gamma}}
        \\
        &-\frac{1}{8R^4}\epsilon_{\dot{\alpha}\dot{\delta}}\epsilon_{\dot{\beta}\dot{\gamma}}x^{\dot{\sigma}\sigma}x^{\dot{\epsilon}\epsilon}h_{\sigma\epsilon,\dot{\sigma}\dot{\epsilon}}+ (\dot\alpha \leftrightarrow \dot\beta)+(\dot\gamma\leftrightarrow\dot\delta)+(\dot\alpha\dot\gamma \leftrightarrow \dot\beta\dot\delta),
    \end{split}
\end{equation}
\begin{equation}
\label{9nov5}
    \begin{split}
        \Bar{F}_{\dot{\alpha}\dot{\beta}\gamma\delta}=&\left(1-\frac{x^2}{4R^2}\right)^2\epsilon^{\alpha\beta}\epsilon^{\dot{\gamma}\dot{\delta}}\frac{\partial}{\partial x^{\dot{\alpha}\alpha}}\frac{\partial}{\partial x^{\dot{\gamma}\gamma}}h_{\beta\delta,\dot{\beta}\dot{\delta}}
-\frac{1}{2R^2}\left(1-\frac{x^2}{4R^2}\right)x^{\dot{\sigma}\sigma}\frac{\partial}{\partial x^{\dot{\alpha}\gamma}}h_{\sigma\delta,\dot{\sigma}\dot{\beta}}
        \\
        &+\frac{1}{4R^2}\left(1-\frac{x^2}{4R^2}\right)\left(-2+x^{\dot{\sigma}\sigma}\frac{\partial}{\partial x^{\dot{\sigma}\sigma}}\right)h_{\delta\gamma,\dot{\alpha}\dot{\beta}}
        + (\dot\alpha \leftrightarrow \dot\beta)+(\gamma\leftrightarrow\delta)+(\dot\alpha\gamma \leftrightarrow \dot\beta\delta)
    \end{split}
\end{equation}
and similarly for complex conjugate components. 
Gauge variation in our coordinates reads
\begin{equation}
\label{9nov6}
    \delta h_{\alpha\dot{\alpha},\beta\dot{\beta}}=-2\left(1-\frac{x^2}{4R^2}\right)\left(\frac{\partial \xi_{\alpha\dot{\alpha}}}{\partial x^{\dot{\beta}\beta}}+\frac{\partial \xi_{\beta\dot{\beta}}}{\partial x^{\dot{\alpha}\alpha}}\right)+\frac{1}{2R^2}\left(2\epsilon_{\alpha\beta}\epsilon_{\dot{\alpha}\dot{\beta}}(x_{\sigma\dot{\sigma}}\xi^{\dot{\sigma}\sigma})-x_{\alpha\dot{\alpha}}\xi_{\beta\dot{\beta}}-x_{\beta\dot{\beta}}\xi_{\alpha\dot{\alpha}}\right).
\end{equation}

With these explicit formulas at hand, we evaluate field strengths using the ansatz (\ref{9nov7}) for the potential.
Then we solve the resulting equations, fixing the residual gauge symmetry as in the spin-$\frac{3}{2}$ case. Eventually, we find 
\begin{equation}
\label{9nov8}
    h_{\alpha\beta,\dot{\alpha}\dot{\beta}}=e^{-i\frac{a}{2}}\left(\frac{8ic-b}{8R^2}-1\right)\frac{\mu_{\alpha}\mu_{\beta}\Bar{\lambda}_{\dot{\alpha}}\Bar{\lambda}_{\dot{\beta}}}{\langle
        \mu\lambda\rangle^2}+ie^{-i\frac{a}{2}}\left(\frac{b-2ac}{4R^2}\right)\frac{\mu_{\alpha}\mu_{\beta}(\Bar{\mu}_{\dot{\alpha}}\Bar{\lambda}_{\dot{\beta}}+\Bar{\mu}_{\dot{\beta}}\Bar{\lambda}_{\dot{\alpha}})}{\langle
        \mu\lambda\rangle^2 \langle\lambda x {\mu}]}.
\end{equation}
Again, we observe that a stronger condition (\ref{28oct5}) is satisfied. Eliminating $\mu$-dependence, we, finally, get (\ref{28oct6}).

\section{Details on Amplitudes from Symmetries}
\label{app:e}

In this appendix we give details on how the deformed momentum conservation condition for the three-point amplitudes  (\ref{29oct20}) is analyzed. 

First, we consider the case of genuine functions (\ref{29oct18}). Evaluating the action of ${\cal P}$, (\ref{23oct9}), on the ansatz (\ref{29oct20}), we find
\begin{equation}
\label{29oct21}
\begin{split}
&\left(D_xf+d_{12,3}\frac{\partial f}{\partial x}+D_zf+d_{31,2}\frac{\partial f}{\partial z}\right)\lambda^1_\alpha\bar{\lambda}^1_{\dot{\alpha}}
+\left(D_xf+d_{12,3}\frac{\partial f}{\partial x}+D_yf+d_{23,1}\frac{\partial f}{\partial y}\right)\lambda^2_\alpha\bar{\lambda}^2_{\dot{\alpha}}\\
&+\left(D_yf+d_{23,1}\frac{\partial f}{\partial y}+D_zf+d_{31,2}\frac{\partial f}{\partial z}\right)\lambda^3_\alpha\bar{\lambda}^3_{\dot{\alpha}}
-\frac{\langle 23 \rangle}{\langle 31 \rangle}\left(z\frac{\partial^2f}{\partial z \partial y}+d_{31,2}\frac{\partial f}{\partial y}\right)\lambda^1_\alpha\bar{\lambda}^2_{\dot{\alpha}}\\
&-\frac{\langle 31 \rangle}{\langle 23 \rangle}\left(y\frac{\partial^2f}{\partial z \partial y}+d_{23,1}\frac{\partial f}{\partial z}\right)\lambda^2_\alpha\bar{\lambda}^1_{\dot{\alpha}}-\frac{\langle 31 \rangle}{\langle 12 \rangle}\left(x\frac{\partial^2f}{\partial x \partial z}+d_{12,3}\frac{\partial f}{\partial z}\right)\lambda^2_\alpha\bar{\lambda}^3_{\dot{\alpha}}\\
&-\frac{\langle 12 \rangle}{\langle 31 \rangle}\left(z\frac{\partial^2f}{\partial z \partial x}+d_{31,2}\frac{\partial f}{\partial x}\right)\lambda^3_\alpha\bar{\lambda}^2_{\dot{\alpha}}-\frac{\langle 12 \rangle}{\langle 23 \rangle}\left(y\frac{\partial^2f}{\partial y \partial x}+d_{23,1}\frac{\partial f}{\partial x}\right)\lambda^3_\alpha\bar{\lambda}^1_{\dot{\alpha}}\\
&-\frac{\langle 23 \rangle}{\langle 12 \rangle}\left(x\frac{\partial^2f}{\partial x \partial y}+d_{12,3}\frac{\partial f}{\partial y}\right)\lambda^1_\alpha\bar{\lambda}^3_{\dot{\alpha}}-R^2(\lambda^1_\alpha\bar{\lambda}^1_{\dot{\alpha}}+\lambda^2_\alpha\bar{\lambda}^2_{\dot{\alpha}}+\lambda^3_\alpha\bar{\lambda}^3_{\dot{\alpha}})
=0,
\end{split}
\end{equation}
where
\begin{equation}
\label{29oct22}
D_x\equiv x\frac{\partial^2}{\partial x^2}+\frac{\partial}{\partial x}.
\end{equation}
This equation has four independent components. To make this manifest, we use
the Schouten identities 
\begin{equation}
\label{29oct23}
\begin{split}
\lambda^3_\alpha &= \frac{\langle 32 \rangle}{\langle 12 \rangle}\lambda^1_\alpha + \frac{\langle 31 \rangle}{\langle 21 \rangle}\lambda^2_\alpha,\\
\bar{\lambda}^3_{\dot{\alpha}} &= \frac{\left[ 32 \right]}{\left[ 12 \right]}\bar{\lambda}^1_{\dot{\alpha}} + \frac{\left[ 31 \right]}{\left[ 21 \right]}\bar{\lambda}^2_{\dot{\alpha}},
\end{split}
\end{equation}
to eliminate $\lambda^3$ and $\bar\lambda^3$. Then, the basis is generated by four structures $\lambda_{\alpha}^1\bar{\lambda}_{\dot{\alpha}}^1$, $\lambda_{\alpha}^1\bar{\lambda}_{\dot{\alpha}}^2$, $\lambda_{\alpha}^2\bar{\lambda}_{\dot{\alpha}}^1$ and $\lambda_{\alpha}^2\bar{\lambda}_{\dot{\alpha}}^2$ and (\ref{29oct21}) requires that the coefficient of each structure vanishes.  As a result, we end up with four single-component equations
\begin{equation} \label{a11}
\begin{split}
&D_xf+\frac{y}{x}D_yf+\bigg(1+\frac{y}{x}\bigg)D_zf+2y\frac{\partial^2f}{\partial x \partial y}\\
&+\left(d_{12,3}+d_{23,1}\right)\left( \frac{\partial f}{\partial x}+\frac{y}{x}\frac{\partial f}{\partial y} \right)+d_{31,2}\bigg(1+\frac{y}{x}\bigg)\frac{\partial f}{\partial z}-R^2\bigg(1+\frac{y}{x}\bigg)f=0
\\
&D_xf+\bigg(1+\frac{z}{x}\bigg)D_yf+\frac{z}{x}D_zf+2z\frac{\partial^2f}{\partial x \partial z}\\
&+\left(d_{12,3}+d_{31,2}\right)\left(\frac{\partial f}{\partial x}+\frac{z}{x} \frac{\partial f}{\partial z}\right)+d_{23,1}\bigg(1+\frac{z}{x}\bigg)\frac{\partial f}{\partial y}-R^2\bigg(1+\frac{z}{x}\bigg)f=0
\\
&\frac{1}{x}D_yf+\frac{1}{x}D_zf+\frac{\partial^2f}{\partial x \partial z}+\frac{\partial^2f}{\partial x \partial y}-\frac{\partial^2f}{\partial y \partial z}
+\frac{d_{31,2}}{z}\frac{\partial f}{\partial x}+\frac{d_{31,2}}{x}\frac{\partial f}{\partial z}\\
&\qquad\qquad\qquad\qquad\qquad\qquad\qquad\qquad\;\;+\bigg(\frac{d_{12,3}}{x}+\frac{d_{23,1}}{x}-\frac{d_{31,2}}{z}\bigg)\frac{\partial f}{\partial y}-\frac{R^2}{x}f=0
\\
&\frac{1}{x}D_yf+\frac{1}{x}D_zf+\frac{\partial^2f}{\partial x \partial z}+\frac{\partial^2f}{\partial x \partial y}-\frac{\partial^2f}{\partial y \partial z}
+\frac{d_{23,1}}{y}\frac{\partial f}{\partial x}+\frac{d_{23,1}}{x}\frac{\partial f}{\partial y}\\
&\qquad\qquad\qquad\qquad\qquad\qquad\qquad\qquad\;\;+\bigg(\frac{d_{12,3}}{x}+\frac{d_{31,2}}{x}-\frac{d_{23,1}}{y}\bigg)\frac{\partial f}{\partial z}-\frac{R^2}{x}f=0.
\end{split}
\end{equation}

Solving these equations may seem a formidable problem. We can, however, use the knowledge gained from  direct computations of amplitudes in particular cases in section \ref{sec:6}. Let us first consider the amplitudes with holomorphic products of spinors in the prefactor, as in our ansatz (\ref{29oct18}).    Then we can anticipate that ${\cal A}_{\rm I}$ and ${\cal A}_{\rm II}$ lead to two independent solution of
   (\ref{a11}).
   For simplicity, we focus on the  domain $p^2>0$, so that $i0$-prescription in (\ref{28oct7}) can be ignored.
We find that, indeed, 
\begin{equation}
\label{29oct24}
f=C_1 \cdot (R^2w)^{-\frac{\sum h+1}{2}}I_{\sum h+1}(2R\sqrt{w})+C_2\cdot (R^2w)^{-\frac{\sum h+1}{2}}K_{\sum h+1},(2R\sqrt{w})
\end{equation}
where
\begin{equation}
\label{29oct25}
w\equiv x+y+z
\end{equation}
solves (\ref{a11}).

Similarly, one can consider candidate amplitudes ${\cal A}_{\rm III}$ and ${\cal A}_{\rm IV}$. Bringing them to the form (\ref{29oct18}) and focusing on $p^2>0$, we find 
\begin{equation}
\label{29oct26}
\begin{split}
f=C_3 x^{-d_{12,3}}y^{-d_{23,1}}z^{-d_{31,2}} \cdot (R^2w)^{\frac{\sum h-1}{2}}I_{\sum h+1}(2R\sqrt{w})\\
+C_4 x^{-d_{12,3}}y^{-d_{23,1}}z^{-d_{31,2}} \cdot (R^2w)^{\frac{\sum h-1}{2}}K_{\sum h+1}(2R\sqrt{w}).
\end{split}
\end{equation}
It is straightforward to see that they also solve (\ref{a11}).

In total, we found four solutions to (\ref{a11}) so far. It is also straightforward to see that these solutions are linearly independent. Now we would like to show that (\ref{a11}) do not have other solutions. 
To see that, we will consider (\ref{a11})  in the neighborhood of some regular point $(x_0,y_0,z_0)$ and count how many integration constants have to be specified, to determine all derivatives of $f$ at a given point from (\ref{a11}).

To start, we note that by combining the last two equations of (\ref{a11}) we can get a first order equation
\begin{equation}
\label{30octx1}
\frac{\partial f}{\partial x}\left(d_{23,1}\frac{1}{y}-d_{31,2}\frac{1}{z} \right)+\frac{\partial f}{\partial z}\left( d_{12,3}\frac{1}{x}-d_{23,1}\frac{1}{y} \right)+\frac{\partial f}{\partial y}\left(d_{31,2}\frac{1}{z}-d_{12,3}\frac{1}{x} \right)=0.
\end{equation}
This, in turn, can be used to eliminate all $z$ derivatives of $f$ in favor of other derivatives. It turns out that after this is done only two equations from (\ref{a11}) are independent. For definiteness, we pick the first two. With $z$-derivatives eliminated, they acquire the form
\begin{equation}
\label{30octx2}
\begin{split}
a_{xx} \partial^2_{xx}f + a_{xy}\partial^2_{xy}f+a_{yy}\partial^2_{yy}f+a_x\partial_xf+a_y\partial_yf+af=0,\\
b_{xx} \partial^2_{xx}f + b_{xy}\partial^2_{xy}f+b_{yy}\partial^2_{yy}f+b_x\partial_xf+b_y\partial_yf+bf=0.
\end{split}
\end{equation}
These equations can be regarded as equations at fixed $z$.
Once these  are solved, $z$-dependence can be reconstructed from (\ref{30octx1}), so we will only focus on $(x,y)$-dependence.

Let us now regard $\{f,\partial_xf,\partial_y f,\partial_{yy}f \}$ as the initial data at $(x_0,y_0,z_0)$. Then, we can use (\ref{30octx2}) to solve for $\partial_{xx}f$ and $\partial_{xy}f$ algebraically in terms of the initial data, so these derivatives are not independent. One can explicitly check, that for genuine   $(x_0,y_0,z_0)$ the matrix of coefficients of  $\partial_{xx}f$ and $\partial_{xy}f$ is non-degenerate, so this is, indeed, possible.

To simplify the analysis, we may focus on solutions of (\ref{a11}) up to linear combinations of four linearly independent solutions in (\ref{29oct24}) and (\ref{29oct26}) that we already know. In particular, by subtracting an appropriate linear combination of the known solutions, we can always achieve 
\begin{equation}
\label{30octx3}
\{f,\partial_xf,\partial_y f,\partial_{yy}f \} = \{0,0,0,0\}
\end{equation}
for new solutions we are looking for. Then (\ref{30octx2}) implies that for solutions with (\ref{30octx3}) satisfied, one also has
\begin{equation}
\label{30octx4}
\{\partial_{xx}f, \partial_{xy}f \} = \{0,0 \}.
\end{equation}

Proceeding further, we consider all consequences of (\ref{30octx2}) obtained by applying one derivative. We get four equations of the form
\begin{equation}
\label{30octx5}
\begin{split}
a_{xx} \partial^3_{xxx}f + a_{xy}\partial^3_{xxy}f+a_{yy}\partial^3_{xyy}f+\dots =0,\\
a_{xx} \partial^3_{yxx}f + a_{xy}\partial^3_{yxy}f+a_{yy}\partial^3_{yyy}f+\dots=0,\\
b_{xx} \partial^3_{xxx}f + b_{xy}\partial^3_{xxy}f+b_{yy}\partial^3_{xyy}f+\dots =0,\\
b_{xx} \partial^3_{yxx}f + b_{xy}\partial^3_{yxy}f+b_{yy}\partial^3_{yyy}f+\dots=0,
\end{split}
\end{equation}
where $\dots$ denotes lower-derivative terms. We find that the matrix of coefficients in front of four highest-derivative terms is non-degenerate for genuine $(x_0,y_0,z_0)$. Hence, considering (\ref{30octx3}), (\ref{30octx4}), we find that all third-order derivatives of $f$ also vanish.

This analysis should be repeated iteratively for higher orders as well. Differentiating (\ref{30octx2}) $n$ times, we obtain $2(n+1)$ equations for $n+3$ derivatives of $f$ of $(n+2)$'th order. So, for $n>1$ the system of equations will be overdetermined. The matrix of coefficients of highest-derivative terms still consists of $a$'s and $b$'s defined by the original equation (\ref{30octx2}). It is not hard to see that it has rank $(n+3)$\footnote{To start, it has four independent rows by virtue of non-degeneracy of the matrix in (\ref{30octx5}). Moreover, a simple inspection shows that one can 
 add to these four rows other $n-1$ rows, so that each time when we are adding a new row it has a non-zero element in a column, in which previously considered raws had vanishing entires. This ensures that the matrix has $n+3$ linearly independent rows.}, so each time highest derivatives of $f$ can be expressed in terms of lower ones and, hence, set to zero. 
 Thus, we find that if (\ref{30octx3}) is imposed, all derivatives of $f$ at $(x_0,y_0,z_0)$ are vanishing. Putting differently,  (\ref{a11}) has only four solutions given in (\ref{29oct24}) and (\ref{29oct26}).

Let us now consider a distributional ansatz (\ref{30oct1}). Substituting it into (\ref{29oct20}) and simplifying we get
\begin{equation}
\label{5nov1}
   [ 12]^{d_{12,3}}[ 23]^{d_{23,1}}[ 31]^{d_{31,2}}\left(p^c+\frac{3+ h}{2R^2}\frac{\partial}{\partial p_c}+\frac{1}{2R^2}p^a\frac{\partial}{\partial p^a}\frac{\partial}{\partial p_c}-\frac{1}{4R^2}p^c\Box_p \right)g\left(\Box_p\right)\delta^{(4)}(p)=0.
\end{equation}
Commuting $p^c$ trough derivatives to the left gives
\begin{equation}
\label{4dec10}
     \left(\left(\Box_p+4R^2\right)g'\left(\Box_p\right)+(1- h)g\left(\Box_p\right)\right)\frac{\partial \delta^{(4)}}{\partial p_c}=0.
\end{equation} 
By requiring the left hand side to be zero, we get a first order differential equation on $g$. Solving it, we obtain the amplitude
\begin{equation}
\label{5nov2}
\begin{split}
    \mathcal{A} &= C_1[  12]^{d_{12,3}}[ 23]^{d_{23,1}}[ 31]^{d_{31,2}}\left(1+\frac{\Box_p}{4R^2}\right)^{h-1}_+\delta^{(4)}(p)\\
    &+C_2[ 12]^{d_{12,3}}[ 23]^{d_{23,1}}[ 31]^{d_{31,2}}\left(1+\frac{\Box_p}{4R^2}\right)^{ h-1}_-\delta^{(4)}(p).
\end{split}
\end{equation}
Analogously, one can find distributional solutions by isolating a prefactor that saturates the homogeneity degrees in spinors required by the helicity constraint with products of undotted spinors.

% BIBLIOGRAPHY
% use BIBTEX if you want
%\bibliographystyle{JHEP}
%\bibliography{yourBIBfiles}

% The bibliography will probably be heavily edited during typesetting.
% We'll parse it and, using the arxiv number or the journal data, will
% query inspire, trying to verify the data (this will probalby spot
% eventual typos) and retrive the document DOI and eventual errata.
% We however suggest to always provide author, title and journal data:
% in short all the informations that clearly identify a document.

\bibliography{shv2}
\bibliographystyle{JHEP}

\end{document}